\documentclass[12pt]{article}

\usepackage[T1]{fontenc}
\usepackage[utf8]{inputenc}
\usepackage{lmodern}
\usepackage{microtype}
\usepackage[margin=1in]{geometry}
\usepackage{setspace}
\usepackage{amsmath,amssymb}
\usepackage{booktabs}
\usepackage{tabularx}
\usepackage{threeparttable}
\usepackage{longtable}
\usepackage{pdflscape}
\usepackage{graphicx}
\usepackage{caption}
\usepackage{subcaption}
\usepackage{tikz}
\usetikzlibrary{arrows.meta,positioning,fit,backgrounds,shapes.geometric}
\usepackage{authblk}
\usepackage{csquotes}
\usepackage{enumitem}
\usepackage{xcolor}
\usepackage{float}
\usepackage[section]{placeins}
\usepackage[hidelinks]{hyperref}
\usepackage[nameinlink,noabbrev]{cleveref}

\usepackage[
  backend=biber,
  style=authoryear-comp,
  maxcitenames=2,
  maxbibnames=99,
  uniquelist=false,
  uniquename=init,
  giveninits=true,
  doi=true,
  url=true,
  isbn=false
]{biblatex}
\newcommand{\sym}[1]{\ifmmode^{#1}\else\(^{#1}\)\fi}

\title{\textbf{Disclosed Human-Capital Disruption and Firm-Specific Risk}}
\author{Ang Zhang}
\affil{Lindner College of Business, University of Cincinnati}
\date{July 2026}

\begin{document}
\maketitle

\begin{abstract}
\noindent
Human capital is a central organizational input, but standard financial data reveal
little about firm-specific disruptions to workforce availability, cost, skills, and
continuity. I construct a measure of disclosed human-capital disruption from earnings
calls using author-defined coding criteria and a contextual language model. Within
firms, a one-standard-deviation increase in the annual
measure is associated with 0.55 percentage points higher idiosyncratic volatility,
0.58 percentage points higher downside deviation, and a 0.46 percentage point lower
worst monthly return, with no corresponding relation to market beta. The results
are stable across seven broader and narrower classification rules and remain after
removing explicit labor-shortage passages and controlling for a recently published
labor-shortage measure and transcript-wide negative and uncertain language.
At the call level, human-capital disruption predicts approximately 0.50\% higher
idiosyncratic volatility over the following 42 trading days after conditioning on
pre-call risk. Risk is elevated before the call as well, and the score predicts the
continuation of that firm-specific risk state over the following 42 trading days.
After leadership and succession passages are removed, the measure also predicts
subsequent named-executive roster exits and the incumbent CEO's effective departure
from office. Earnings calls therefore reveal disturbances to a key organizational
input that are broader than labor shortages and informative about the distribution of
firm outcomes.
\end{abstract}



\newpage
\section{Introduction}
\label{sec:introduction}

Human capital is a productive input, a repository of firm-specific knowledge, and an
important component of organizational capital. Yet financial statements reveal little
about when that input becomes difficult to acquire, more costly to retain, unavailable
for production, or disrupted by organizational change. Headcount and labor expense
record quantities after the fact. They do not reveal whether a firm is struggling to
hire skilled workers, absorbing unusually high turnover, responding to wage pressure,
reorganizing its workforce, or managing a disruption to leadership and continuity.
These conditions can constrain production, increase adjustment costs, and widen the
distribution of firm outcomes even when management ultimately resolves them.

This paper asks whether managers' discussion of human-capital disruption reveals a
firm-specific risk state. I define human-capital disruption as a material condition or
action that constrains, reprices, interrupts, or reorganizes the reporting firm's
workforce. The definition includes labor shortages, skill constraints, attrition and
retention pressure, wage and labor-cost pressure, safety and continuity problems,
workforce restructuring, and consequential leadership transitions. It excludes
routine headcount disclosures, generic statements about culture or employees,
workforce conditions at other firms, and unsupported analyst questions. The measure
is broader than labor-shortage exposure but narrower than general workforce
discussion.

The economic prediction follows from viewing human capital as an organizational
input. Organization capital commands a risk premium
\parencite{eisfeldt2013organization}; intangible capital affects investment and firm
value \parencite{peters2017intangible}; and employee satisfaction, mobility, and
turnover are related to valuation and performance
\parencite{edmans2011stock,campbell2012who,li2022employee}. When access to needed labor
becomes uncertain, retention becomes unusually costly, or a firm undertakes a major
workforce adjustment, its ability to execute a given operating plan becomes less
certain. The resulting risk should appear most clearly in firm-specific volatility,
downside dispersion, and tail outcomes, particularly when labor is an important
production input.

I measure these conditions in earnings calls. Calls occur frequently, combine
prepared remarks with analyst questions, and allow managers to explain operating
conditions with no dedicated financial-statement line item
\parencite{matsumoto2011conference}. The sample contains 45,725 calls linked to CRSP
and Compustat from October 2005 through May 2025. I use 50 excerpts that I classified
to define the economic boundary and written coding criteria, then apply those criteria
at scale with a contextual language model. The measure counts disruption excerpts,
normalizes by transcript length, and aggregates calls within Compustat fiscal years.
The regressions also control for the frequency of broad workforce-related language.

The annual evidence supports the organizational-risk interpretation. The primary
sample contains 13,201 call-covered firm-years from 2,870 firms over fiscal years
2006--2024. In regressions with firm and year fixed effects, a
one-standard-deviation increase in human-capital disruption is associated with 0.55
percentage points higher annualized idiosyncratic volatility, 0.58 percentage points
higher downside deviation, and a 0.46 percentage point more negative worst monthly
return. Market beta is unrelated to the measure. Disruption is also more prevalent in
labor-intensive industries, and its relation with idiosyncratic volatility is stronger
in those industries.

The economic results are stable across alternative readings of borderline excerpts.
Seven broader and narrower classification rules change the number of accepted
excerpts by between \(-37\%\) and \(+72\%\), yet all seven preserve the signs and
statistical significance of the three principal return-distribution results.
Firm-year ranks remain highly correlated across the alternative rules. This stability
matters for a construct that combines several forms of disruption under a common
organizational-input interpretation.

The recently published labor-shortage measure of \textcite{harford2026labor} provides
the closest economic comparison. Labor shortages capture difficulty obtaining an
essential input. Human-capital disruption additionally captures retention, cost,
skills, restructuring, safety, continuity, and leadership disturbances. In the common
annual sample, the disruption coefficient is 0.65 percentage point
(\(t=2.97\)) after controlling jointly for the released labor-shortage measure and
transcript-wide negative and uncertain language. Removing explicit shortage passages
leaves a coefficient of 0.68 percentage point (\(t=3.28\)). The result therefore
extends beyond measured labor shortages and generic adverse tone.

Call timing provides complementary evidence. Conditional on the nearest pre-call risk
realization and information available before the call, disruption predicts higher
idiosyncratic volatility in each of the first two nonoverlapping 21-trading-day
blocks. Over the following 42 trading days, a one-standard-deviation increase in
disruption predicts approximately 0.50\% higher idiosyncratic volatility. Risk is also
elevated before calls with greater disruption, while post-minus-pre changes are small.
Earnings-call language thus reveals a persistent organizational condition and contains
information about its near-term continuation.

Organizational outcomes reinforce this interpretation. After removing every accepted
excerpt that mentions leadership or succession, the measure predicts next-year
named-executive roster exits and the incumbent CEO's effective departure from office
before the next observed call. The CEO result remains positive after excluding
intervals with evidence of a prior transition disclosure in Item 5.02 filings. These
outcomes connect the signal to subsequent changes inside the organization without
using direct transition language to construct the score.

The paper contributes to research on human and organization capital by measuring
disturbances to the workforce as a productive input and connecting them to the
distribution of firm-specific outcomes
\parencite{eisfeldt2013organization,peters2017intangible,donangelo2019labor}. It also
adds a distinct construct to the firm-level text-measure literature
\parencite{hassan2019firm,sautner2023firm,florackis2023cyber,wu2024text}. The direct
comparison with \textcite{harford2026labor} locates labor-shortage exposure within a
broader set of workforce disturbances and shows that the broader measure carries
incremental firm-specific risk information. Finally, the call-timed and organizational
results show that earnings-call language contains information about the persistence of
firm-specific risk and subsequent organizational adjustment.

The remainder of the paper develops the economic framework, describes the sample and
measurement procedure, presents the validation evidence, and reports the annual,
call-timed, and organizational results.

\section{Related Literature and Economic Framework}
\label{sec:literature}

\subsection{Firm-level measures from text}

Text allows researchers to measure firm-specific conditions that do not appear in
structured financial data. Early finance applications show that corporate language
contains information not summarized by accounting quantities
\parencite{tetlock2008more,loughran2011liability}. The broader text-as-data literature
places construct definition, text selection, and aggregation within the economic
design of the study \parencite{gentzkow2019text}.

\textcite{hassan2019firm} provide the closest methodological template. They measure
firm-level political risk from earnings calls by conditioning political discussion on
risk language, normalize the resulting count by transcript length, and validate the
measure with excerpt review, aggregate patterns, falsifications, corporate actions,
and political behavior. Their design separates topic exposure, risk conditional on
the topic, and economic outcomes. The current paper follows that logic by retaining a
broad workforce-discussion control and defining the focal variable as the frequency of
excerpts that convey a firm-specific disruption.

Recent firm-level measures connect text to distinct economic manifestations of the
underlying construct. \textcite{sautner2023firm} combine human audits, score
perturbations, industry patterns, green employment and innovation, and option-market
outcomes for climate exposure. \textcite{florackis2023cyber} connect cybersecurity
language to future attacks and asset prices. \textcite{wu2024text} relate
supply-chain-risk exposure to realized and option-implied volatility, corporate
policies, and a topic-specific shock. These papers motivate a validation strategy that
combines content evidence, known groups, alternative measurement choices, external
outcomes, and economic predictions.

Contextual language models expand the set of distinctions that can be measured in
corporate text. BERT and DeBERTa represent negation, word sense, and contextual
dependencies that bag-of-words procedures miss
\parencite{devlin2019bert,he2021deberta}. \textcite{jha2026finance} use contextual
embeddings and extensive internal and external checks to measure historical sentiment
toward finance. \textcite{becht2026benefits} combine prompted language-model analysis,
traditional natural-language processing, and human labeling in a top-finance
application. The present design uses author-defined criteria to establish the economic
boundary, a contextual classifier to scale those criteria, and aggregate economic
evidence to evaluate the resulting firm-level measure.

\subsection{Human capital, organization capital, and firm risk}

Human capital affects firm value through production, knowledge, coordination, and
adjustment. \textcite{eisfeldt2013organization} show that organization capital is
priced, while \textcite{peters2017intangible} demonstrate that recognizing intangible
capital changes investment measurement. Employee satisfaction is associated with
equity value \parencite{edmans2011stock}, and the departure or mobility of skilled
employees can impose costs on source firms \parencite{campbell2012who}. Concentrated
key human capital is related to volatility and exposure to employee departures
\parencite{israelsen2017key}. These results motivate an organizational-input view:
workforce conditions matter not only because labor is an expense, but also because
people embody knowledge and production capacity that may be costly to replace.

The risk can operate without an immediate decline in average performance. A shortage
may constrain output, attrition may weaken continuity, and wage pressure may reprice
the input. A restructuring may improve future efficiency while creating execution
risk today. The common feature is reduced operating flexibility and greater
uncertainty about how the firm will deliver a given plan. Labor leverage provides one
mechanism by which labor-intensive production affects equity risk
\parencite{donangelo2019labor}. Labor-market insurance considerations also shape
corporate financing decisions \parencite{agrawal2013labor}. These mechanisms predict
that a disruption signal may load more strongly on firm-specific volatility and
downside states than on market beta or a uniformly lower conditional mean.

Workforce turnover supplies an organizational outcome with a different measurement
error. Large-sample evidence links employee turnover to firm performance
\parencite{li2022employee}, and current work develops text measures of human-capital
disclosure and workforce turnover
\parencite{demers2024measuring,dey2025human}. CEO transitions are economically
important, but their interpretation requires care. Managerial styles affect policy
choices, shocks to CEO availability affect firms, and performance can induce turnover
\parencite{bennedsen2020do,jenter2021performance}. Relative performance also enters
turnover decisions \parencite{jenter2015ceo}. In this paper, annual roster absence and
effective CEO-office exit provide organizational outcomes distinct from stock-return
risk. The principal tests remove direct leadership and succession language from the
human-capital score.

The SEC's 2020 modernization of Regulation S-K likewise recognizes human-capital
resources and measures as potentially material to investors
\parencite{sec2020modernization}.

\subsection{Labor shortages and human-capital disruption}

\textcite{harford2026labor} fine-tune FinBERT to detect labor-shortage sentences and
aggregate the results into firm-level exposure. Their validation uses state and
industry labor-market conditions, staff expense, immigration policy, and federal
unemployment programs. They examine announcement returns, subsequent market and
operating outcomes, labor--capital substitution, and process and artificial-intelligence
innovation. It provides the closest economic comparison for this study.

A labor shortage is a disturbance to workforce availability. The construct in
this paper also includes disruptions that occur after labor has been hired: attrition
and retention pressure, wage repricing, skill constraints, workforce adjustment,
safety and continuity problems, and consequential leadership transitions. These
conditions share a common link to the firm's ability and cost of deploying human
capital, while implying different operating responses. The classification retains only
material firm-specific disturbances or consequential adjustments; routine workforce
metrics and general employee discussion remain outside the construct.

This distinction yields observable implications. Explicit shortage passages should
account for part, but not all, of the disruption score. The broader measure should
retain firm-specific risk information after the HHQ measure and general adverse tone
are included, and after direct shortage language is removed. These implications
motivate two empirical comparisons. Section~\ref{sec:validation} examines overlap
with labor-shortage exposure, and Section~\ref{sec:baseline} tests whether the broader
measure retains incremental economic content.

\subsection{Economic implications}

The framework motivates four empirical predictions.

First, disclosed disruption should be positively related to firm-specific risk.
Idiosyncratic volatility captures variation not explained by common Fama--French
factors \parencite{fama1993common,campbell2001idiosyncratic}. Downside deviation and
worst-month returns ask whether the relation is concentrated in adverse states, a
distinction related to the broader downside-risk literature \parencite{ang2006downside}.
If the signal primarily reflects conditions specific to the firm's workforce and
organization, it should have little relation to market beta.

Second, the relation should be stronger when labor is a more important production
input. Labor-intensive firms have less room to absorb availability and cost shocks
without changing output, staffing, or margins. This cross-sectional implication
provides a construct-validity test of the organizational-input interpretation.

Third, the disruption signal should contain information beyond broad workforce
salience and explicit shortage language. The empirical tests condition on the amount
of workforce discussion, delete explicit shortage passages, and include the released
\textcite{harford2026labor} measure.

Fourth, the timing of the relation can distinguish a persistent organizational
condition from a disclosure-date shock. I compare risk before the call with
conditional risk afterward and with post-minus-pre changes. A positive relation on
both sides of the call without a discontinuity would characterize persistence rather
than the arrival of a new risk state.

The empirical analysis follows this sequence. Annual return-distribution regressions
establish contemporaneous economic content. Call-timed regressions test near-term
predictive content, while executive turnover and CEO-office exit provide
organizational outcomes measured outside the text.

\section{Data and Sample Construction}
\label{sec:data}

\subsection{Earnings-call corpus}

I combine two publicly released transcript archives. The first contains 18,755
earnings-call transcripts scraped from The Motley Fool over 2017--2023, concentrated
in 2019--2022 \parencite{tpotterer2023motley}. The second contains 33,362 calls for
S\&P 500 constituents and other large U.S. firms from 2005--2025
\parencite{kurry2025sp500}. After excluding transcripts shorter than 500 characters
and retaining the longer record when both sources contain the same ticker, date, and
quarter, the combined corpus contains 49,968 calls dated from October 13, 2005 through
May 15, 2025. I match ticker and call date to CRSP security-name histories and then to
Compustat through the CRSP/Compustat Merged link history. Date-specific matching
prevents a reused ticker from being assigned to the wrong firm.

Some earnings calls appear more than once because source metadata or ticker aliases
differ. When duplicate records share a ticker and call date, I retain the transcript
with the largest word count. I then retain one record for each linked firm and call
date, prioritizing a date-valid ticker link over a unique-ticker fallback and then the
longer transcript. This final step removes 26 alias rows across 26 calls. The resulting
sample contains one transcript for each of 45,725 linked earnings calls. Each observation records the
transcript source, ticker, call date, transcript length, fiscal quarter, and fiscal
year.

The annual measure is aligned to Compustat fiscal years rather than calendar years. For
firm \(i\) and fiscal year \(t\), calls in
\((Datadate_{i,t-1},Datadate_{i,t}]\) enter the numerator and denominator for year
\(t\). Counts and transcript words are summed across all retained calls in that
interval. This rule assigns disclosure to the fiscal period in which it occurs and
avoids equating a December calendar-year observation with a non-December fiscal year.
The annual analysis covers fiscal years 2006--2024; the partial 2005 and 2025
boundaries are excluded.

\subsection{Compustat, CRSP, and Fama--French data}

The initial Compustat sample contains 302,028 industrial, consolidated, domestic,
standard-format firm-years. Compustat supplies fiscal dates,
identifiers, assets, sales, employment, profitability, capital expenditure, R\&D,
debt, cash, equity, and industry codes. The linked panel has 13,705 firm-years
with at least one call and 2,872 firms. Restricting to the complete FY2006--FY2024
boundary yields 13,201 firm-years and 2,870 firms.

Security links are selected separately for call dates and fiscal endpoints. Call-level
tests use the CRSP security active on the call date. Annual returns use the CCM link
active on the Compustat fiscal endpoint, prioritizing primary links and then standard
link types. This distinction matters when a firm changes its traded security during a
fiscal year. Ten call-covered firm-years have no valid CCM link at the fiscal endpoint
and retain missing annual returns rather than borrowing the security associated with a
different call.

Daily and monthly returns come from CRSP through December 31, 2025. Daily
Fama--French market, size, value, and risk-free factors are matched by trading date
\parencite{fama1993common}.

Annual idiosyncratic volatility is estimated over the 365 calendar days ending at the
fiscal statement date. Daily excess returns are regressed on an intercept and the
three Fama--French factors. The standard deviation of the residuals is annualized by
\(\sqrt{252}\). A firm-year requires at least 60 daily observations and a last return
within ten calendar days of the fiscal endpoint. Market beta is the coefficient on the
market factor from the same regression. The FF3 systematic share is the regression
\(R^2\).

Downside deviation and worst-month returns are constructed from fiscal-year-aligned
return histories. Downside deviation is the annualized dispersion of negative
firm-specific returns. The worst-month variable is the minimum monthly return during
the fiscal window. The fiscal-year total return requires at least ten monthly
observations, at least 300 days between the first and last observation, and no gap
greater than 45 days at either endpoint. These rules prevent a later fiscal year from
inheriting stale returns merely because six earlier monthly observations remain
available.

\subsection{Call-timed return windows}

The call-level analysis uses calls dated no later than December 31, 2024. This cutoff
allows the 2025 CRSP extension to complete future return windows without treating the
partial 2025 transcript corpus as a complete disclosure sample. After the date cap and
alias removal, 44,799 calls enter the forward-risk sample and 44,798 link to daily CRSP
returns. The event day is the first CRSP trading day on or after the recorded call
date. The maximum realized gap is four calendar days.

The immediate response interval, trading days 0 and +1, is excluded from the risk
outcome. The nearest pre-call block ends at day \(-2\), and the first post-call block
begins at day \(+2\). For a \(K\)-trading-day design, returns are collected separately
on each side of the omitted interval and residualized against the same three daily
factors. I examine two nonoverlapping 21-day post-call blocks and
a cumulative 42-day window. A 63-day window measures longer persistence. The common
63-day sample has 41,777 calls and 2,527 firms. The sixteen-block dynamic sample has
41,559 calls and 2,510 firms.

For each call, the next distinct call date within the same gvkey is recorded before
estimating the horizon tests. Among calls with an observed successor call, 97.9\% of
valid 42-day windows end before it, compared with 30.3\% of 63-day windows. A stricter
42-day sample requires an observed successor call and a completed risk window before
that date. It contains 38,357 calls and 2,389 firms.

Accounting controls in call-level models are selected using public information dates.
For each annual Compustat record, the matched fiscal-Q4 report date from Compustat
Quarterly must occur within 180 days after fiscal period end. The latest report date
strictly before the call is selected, and the associated fiscal statement can be no
more than 550 days old. Same-day and future report dates are excluded.

\subsection{Execucomp outcomes}

The annual organizational tests use Execucomp named-executive rosters. For each
firm-year, the roster is compared with the exact next fiscal year when that year is
consecutive. The named-executive roster-exit rate is the fraction of current roster
members absent from the next roster. An indicator records whether any named executive
exits the roster. A CEO roster-absence outcome records whether the named CEO is absent
in the next fiscal year. These are disclosure-based roster outcomes, not exact
company-departure dates. A person can leave the named group without leaving the firm.

The exact-date CEO test uses Execucomp CEO spells. A spell begins at
\texttt{BECAMECEO} and ends at \texttt{LEFTOFC}; the latter is the effective date the
individual leaves the CEO office, not a general officer-departure or announcement
date. Calls from 2008--2024 are ordered within gvkey, and consecutive-call intervals
are retained when the next call occurs 45--150 calendar days later. The risk set
requires exactly one incumbent CEO on the call date and evidence that the firm's
Execucomp coverage continues through the interval endpoint. Missing future coverage is
censoring, not a zero.

The primary event equals one when the incumbent's effective CEO-office exit occurs
strictly after the focal call and strictly before the next call. The fully controlled
sample contains 27,691 nonoverlapping intervals, 1,411 firms, and 715 exits. Controls
include interval length, transcript source, broad workforce salience, public pre-call
accounting information, CEO tenure and age, and trailing return and idiosyncratic
volatility. The nonlinear models treat interval duration as exposure.

\subsection{Labor-shortage comparison data}

I use the call-quarter and firm-year labor-shortage measures released by Harford, He,
and Qiu (HHQ) \parencite{harford2026labor} rather than reconstructing their measure
from the training labels. Their annual file contains 39,291 firm-years with nonmissing
gvkeys, 29.7\% of which are positive. The call-quarter file contains 138,159
observations, 13.1\% of which are positive.

For direct overlap, the released calendar-year variable is matched to the current
measure after aggregating the current calls by gvkey and calendar year. The common
overlap sample contains 8,047 observations. For the joint annual regressions, I match
the HHQ calendar-year observation to the current fiscal-year observation carrying the
same numerical year. Outcome-specific complete-case samples contain 6,986--7,019
observations and 1,206--1,216 firms over 2005--2021. The match is exact for
December-year-end firms but can compare nonidentical reporting intervals for firms
with other fiscal year-ends; I use it only for a common-sample comparison and not as a
replication of HHQ's outcome design.

\begin{table}[!htbp]
\centering
\caption{Sample Construction and Summary Statistics}
\label{tab:sample}
\footnotesize
\begin{threeparttable}
\begin{tabularx}{\linewidth}{Xrr}
\toprule
\multicolumn{3}{l}{\textit{Panel A. Sample construction}}\\
Sample & Observations & Firms\\
\midrule
Initial Compustat sample & 302,028 & \\
Firm-years with at least one linked call & 13,705 & 2,872\\
Main annual sample, FY2006--FY2024 & 13,201 & 2,870\\
Annual idiosyncratic-volatility regression & 12,114 & 2,315\\
Annual downside/worst-month regressions & 11,851 & 2,209\\
Harford--He--Qiu (HHQ) annual overlap sample & 8,047 & \\
HHQ joint annual regressions & 6,986--7,019 & 1,206--1,216\\
Calls in the forward-risk sample & 44,799 & \\
Common 63-trading-day call sample & 41,777 & 2,527\\
Common dynamic-path call sample & 41,559 & 2,510\\
CEO-office-exit analysis intervals & 27,691 & 1,411\\
\end{tabularx}
\medskip
\begin{tabularx}{\linewidth}{Xrrrrrr}
\toprule
\multicolumn{7}{l}{\textit{Panel B. Primary annual sample}}\\
Variable & $N$ & Mean & SD & P25 & Median & P75\\
\midrule
Human-capital disruption frequency & 13,201 & 1.173 & 2.035 & 0.000 & 0.446 & 1.432\\
Any human-capital disruption & 13,201 & 0.618 & 0.486 & 0.000 & 1.000 & 1.000\\
Broad workforce-discussion frequency & 13,201 & 46.449 & 21.416 & 29.286 & 42.882 & 60.066\\
Share of screened workforce excerpts classified as disruptions & 13,201 & 0.025 & 0.038 & 0.000 & 0.011 & 0.034\\
Idiosyncratic volatility & 13,182 & 0.318 & 0.209 & 0.184 & 0.256 & 0.384\\
Market beta & 13,182 & 1.012 & 0.385 & 0.770 & 1.001 & 1.244\\
Fiscal-year return & 13,003 & 0.128 & 0.538 & -0.136 & 0.091 & 0.317\\
Log assets & 13,189 & 9.113 & 1.818 & 7.952 & 9.139 & 10.292\\
Tobin's q & 13,168 & 2.322 & 2.078 & 1.182 & 1.642 & 2.620\\
ROA & 12,563 & 0.097 & 0.154 & 0.054 & 0.106 & 0.161\\
Book leverage & 13,189 & 0.305 & 0.229 & 0.136 & 0.284 & 0.428\\
Cash/assets & 13,189 & 0.156 & 0.181 & 0.035 & 0.089 & 0.203\\
\midrule
\multicolumn{7}{l}{\textit{Panel C. Within- and between-firm variation}}\\
Variable & \multicolumn{2}{c}{Overall SD} & \multicolumn{2}{c}{Within-firm SD} & \multicolumn{2}{c}{Between-firm SD}\\
Human-capital disruption frequency & \multicolumn{2}{c}{2.035} & \multicolumn{2}{c}{1.343} & \multicolumn{2}{c}{1.971}\\
Broad workforce-discussion frequency & \multicolumn{2}{c}{21.416} & \multicolumn{2}{c}{11.198} & \multicolumn{2}{c}{19.298}\\
\bottomrule
\end{tabularx}
\begin{tablenotes}[flushleft]\footnotesize
\item Notes: The main annual sample contains call-covered firm-years from FY2006 through FY2024. Text frequencies are excerpts classified as human-capital disruptions or broad workforce discussion per 10,000 transcript words. Return variables are in decimal units. Observation counts vary with outcome availability. Panel C decomposes the unbalanced panel descriptively; the within-firm SD is the standard deviation after removing the firm mean.
\end{tablenotes}
\end{threeparttable}
\end{table}

\subsection{Controls and estimation samples}

Annual models control for broad workforce-exposure frequency, log assets, Tobin's q,
ROA, book leverage, and cash scaled by assets. Ratio variables and return-distribution
outcomes are winsorized at the first and ninety-ninth percentiles. The focal measure
is standardized within each
sample. The main annual models include firm and fiscal-year fixed effects and cluster
standard errors by firm. Two-way firm and year clustering is reported for the turnover
tests and selected sensitivities, following the general finance-panel guidance in
\textcite{petersen2009estimating,cameron2011robust}.

Table~\ref{tab:sample} reports the sample flow and descriptive statistics. The measure
is positive in 61.8\% of the main-sample firm-years, with a mean of 1.17 accepted
excerpts per 10,000 transcript words. Its within-firm standard deviation is 1.34, which
is substantial relative to the overall standard deviation of 2.04. The fixed-effect
tests therefore do not rely only on persistent cross-sectional differences in which
firms discuss their workforce.

\section{Measuring Human-Capital Disruption}
\label{sec:measure}

\subsection{Economic construct}

An excerpt is classified as a human-capital disruption when it describes a material,
firm-specific disturbance to human capital. The
disturbance can arise from acquisition, retention, cost, productivity, availability,
safety, workforce organization, or leadership. It can appear as an adverse condition,
such as acute staffing difficulty or attrition pressure; as a consequential action,
such as a workforce reduction or redeployment; or as forward-looking uncertainty
about the firm's ability to obtain or organize needed labor. The common requirement is
that the reporting firm's own human-capital input is constrained, repriced, disrupted,
or materially reorganized.

The classification is not a sentiment measure. A firm may describe a costly workforce
adjustment positively because management expects the action to improve future
performance. The excerpt remains positive if it documents a material disruption or
consequential adjustment. By contrast, a statement that the firm has no staffing
problem, attrition is low, or labor pressure has disappeared is negative. The
distinction separates evidence that a disruption or adjustment occurred from
management's assessment of its resolution.

Four recurring types of workforce discussion are excluded. First, routine metrics such
as headcount growth or the number of employees do not establish disruption. Second,
generic statements about culture, diversity, training, or employees as an asset are
workforce discussion but not evidence of a disturbance. Third, the workforce must
belong to the reporting firm; a customer's, supplier's, or economy-wide labor
condition qualifies only when the call identifies a material consequence for the
reporting firm. Fourth, an analyst's unsupported question is insufficient when neither
the question nor management's response identifies a material condition.

The accepted excerpts capture several manifestations of the same construct. Shortages,
attrition, wage pressure, restructuring, safety, and leadership
change can affect the firm through different channels. Their common element is a
material disturbance to an organizational input that can change operating flexibility
and the distribution of firm outcomes. I retain a binary excerpt classification
because separately coding severity, tone, resolution, and mechanism would introduce
additional judgments that require their own validation.

Because the measure begins with earnings-call language, it captures disruption that
management discusses or analysts elicit in an investor-facing setting. It does not
measure unspoken workforce conditions or employee experience outside that disclosure
channel.

\subsection{Text screening and context}

I begin with a vocabulary of workforce-related terms. The vocabulary identifies
1,835,450 sentences in the combined transcript corpus. I then use
\texttt{all-mpnet-base-v2} sentence representations and a radial-basis support-vector
classifier trained on 3,110 Opus-labeled sentences to distinguish substantive
human-capital discussion from incidental word use. A threshold of 0.65, selected on a
separate 400-sentence review sample, yields precision of 0.922 and recall of 0.769 for
this broad screening task. The screen retains 173,273 sentences across 36,159 calls
for the more demanding disruption classification. This two-stage design reduces
obviously irrelevant text but also defines the scope of the measure: relevant
sentences omitted by the vocabulary or the first-stage classifier cannot be recovered
later. Section~\ref{sec:validation} and the Appendix quantify this limitation.

The classification unit is a candidate sentence together with the immediately
preceding and following sentences. The accepted count is attributed only to the
candidate sentence, so overlapping context does not create multiple observations.
The context helps determine whether ``turnover'' refers to employees or inventory,
whether a constraint belongs to the reporting firm, whether a statement negates a
problem, and whether management answers an analyst's question.

\subsection{Author-guided classification}

I manually classified 50 excerpts drawn from the first-stage candidate pool and
stratified across its classification scores. The exercise clarified the boundary
between general workforce discussion and material disruption and supplied examples
for the written inclusion and exclusion criteria. These 50 excerpts did not enter the
subsequent DeBERTa training or evaluation samples. Appendix
Table~\ref{tab:app_examples} presents representative inclusions, exclusions, and
boundary cases.

Claude Opus 4.8 applied the same criteria to a training sample of 729 excerpts, of
which 274 were classified as disruptions, and to a separate 400-excerpt evaluation
sample containing 219 positives. I then fine-tuned three
\texttt{microsoft/deberta-v3-base} models on the 729 labeled excerpts and averaged
their decision scores. DeBERTa is a contextual transformer that can represent
negation and word meaning jointly across the three-sentence input
\parencite{he2021deberta}. Using three independently initialized models also permits
alternative measures that require agreement across model runs.

On the 400-excerpt threshold-selection sample, the primary threshold is chosen to
achieve at least 90\% precision relative to the Opus labels. A lower threshold that
maximizes \(F_1\) provides an inclusive alternative. The classifier output is used
only to accept or reject excerpts; the firm-level measure is constructed from
accepted counts rather than interpreted as a probability. Appendix
Table~\ref{tab:app_operating} reports the model-level statistics, thresholds, and
sentence-level precision and recall. Section~\ref{sec:validation} reports the annual
aggregation audit and the author comparison separately.

\subsection{Call and firm-year measures}

For call \(c\), the primary measure is
\begin{equation}
\label{eq:call_measure}
  HCDisruption_{c}
  =
  10{,}000
  \frac{\sum_{s\in c}\mathbf{1}\{\widehat p_s\geq T_{P90}\}}
       {Words_c},
\end{equation}
where \(\widehat p_s\) is the ensemble decision score for excerpt \(s\).
The scale gives the number of accepted excerpts per 10,000 transcript words. I also
construct a broad workforce-discussion measure,
\begin{equation}
\label{eq:exposure_measure}
  HCExposure_{c}
  =
  10{,}000
  \frac{\#\{\text{workforce-vocabulary sentences in }c\}}
       {Words_c}.
\end{equation}
The broad measure is determined before the disruption classification. Including it as
a control asks whether the content of workforce discussion matters beyond how much
the call discusses employees.

Call counts and transcript words are summed within each Compustat fiscal-year window:
\begin{equation}
\label{eq:annual_measure}
  HCDisruption_{i,t}
  =
  10{,}000
  \frac{\sum_{c\in(i,t)}\sum_{s\in c}
       \mathbf{1}\{\widehat p_s\geq T_{P90}\}}
       {\sum_{c\in(i,t)} Words_c}.
\end{equation}
This construction weights calls by transcript length rather than averaging
call-specific ratios. Alternative measures use the raw accepted count, the accepted
share among workforce sentences, an indicator for any accepted excerpt, and the
inclusive maximum-\(F_1\) threshold. Regression specifications standardize the
frequency within the relevant estimation sample.

\subsection{Robustness to alternative classification rules}

I reconstruct the measure under seven rules that vary the confidence threshold,
require agreement across independently trained classifiers, or exclude explicit
labor-shortage and leadership passages. Section~\ref{sec:validation} compares their
firm-year rankings and economic results; Appendix
Tables~\ref{tab:app_variant_size}--\ref{tab:app_economic_grid} report the complete
counts and coefficient grid.

\section{Validation and Construct Differentiation}
\label{sec:validation}

The validation evidence asks whether the measure behaves like a disturbance to an
organizational input, remains economically informative under reasonable alternative
classifications, extends beyond labor shortages and generic adverse language, and
retains the author-defined economic boundary when applied at scale.
Table~\ref{tab:validation} summarizes these complementary tests.

\begin{table}[!htbp]
\centering
\caption{Economic and Measurement Validation of Human-Capital Disruption}
\label{tab:validation}
\footnotesize
\begin{threeparttable}
\begin{tabularx}{\linewidth}{>{\raggedright\arraybackslash}p{3.0cm}>{\raggedright\arraybackslash}p{4.2cm}>{\raggedright\arraybackslash}X}
\toprule
Validity dimension & Design & Principal evidence\\
\midrule
Cross-sectional construct validity & Disruption frequency and its risk relation across industries with high versus low labor intensity & Mean score 1.78 versus 0.92; disruption $\times$ high-labor-industry coefficient=0.0085 ($t=2.88$).\\[3pt]
Economic stability to classification ambiguity & Seven broader and narrower classifications applied to the same firm-year samples and specifications & Accepted excerpts -37\% to +72\%; Spearman correlations 0.889--0.989; minimum $|t|$=3.16, 3.73, and 3.46 for the three principal outcomes.\\[3pt]
Incremental content beyond labor shortages and tone & Annual regression with HHQ labor-shortage exposure and transcript-wide negative and uncertainty frequencies & Disruption coefficient=0.0065 ($t=2.97$); after deleting explicit shortage passages, 0.0068 ($t=3.28$).\\[3pt]
Organizational convergence & CEO-office exit before the next call, using the score after removing leadership and succession passages & Coefficient=0.391 percentage points ($t=2.53$), or 15.1\% of the event rate; next-year named-executive roster outcomes are also positive.\\[3pt]
Author guidance and model fidelity & Fifty author-classified examples define the criteria; the final classifier is compared with these examples & Author-example agreement=0.76 ($\kappa=0.519$); precision=0.75 and recall=0.75.\\[3pt]
Agreement on difficult excerpts & Independent author and Opus classifications of 120 excerpts concentrated near the disruption boundary & Agreement=0.65; $\kappa=0.300$ (95\% interval 0.134--0.464).\\[3pt]
Aggregation fidelity & Stratified review of 100 ticker--calendar-year observations, conditional on the workforce candidate screen & Population-weighted precision=1.000 and recall=0.918 for a binary any-disruption indicator.\\[3pt]
\bottomrule
\end{tabularx}
\begin{tablenotes}[flushleft]\footnotesize
\item Notes: The three minimum absolute $t$-statistics in the ambiguity row correspond to idiosyncratic volatility, downside deviation, and worst monthly return. The HHQ comparison uses the common annual sample and includes firm and year fixed effects, broad workforce discussion, and standard firm controls. The CEO-office-exit model also includes transcript source, pre-call accounting information, CEO characteristics, interval length, trailing return, and trailing idiosyncratic volatility. The aggregation audit evaluates a binary any-disruption indicator, rather than the continuous gvkey--fiscal-year score used in the regressions. Its 100 ticker--calendar-year observations are drawn conditional on the workforce candidate screen and weighted to the population shares of the classifier-positive and classifier-negative strata.
\end{tablenotes}
\end{threeparttable}
\end{table}

\subsection{Cross-sectional construct validity}

\begin{figure}[!htbp]
\centering
\includegraphics[width=\linewidth]{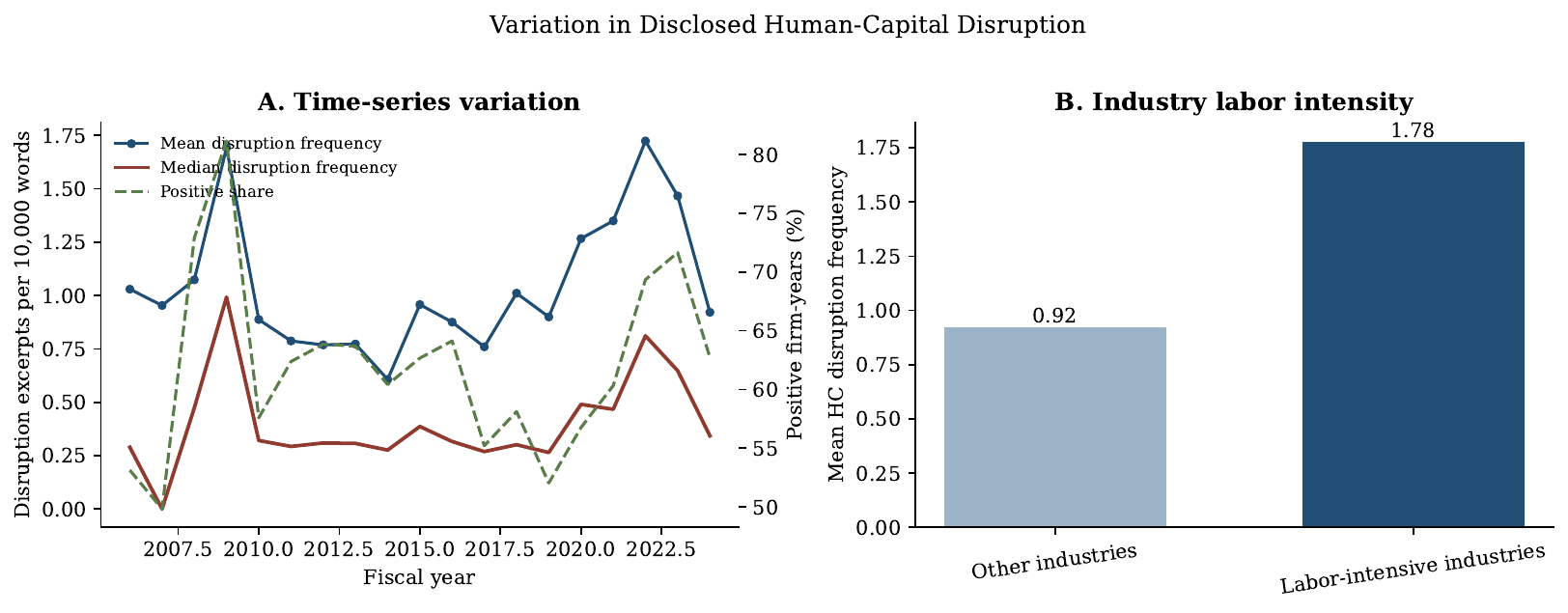}
\caption{Variation in Disclosed Human-Capital Disruption}
\label{fig:known_groups}
\begin{minipage}{0.94\linewidth}
\footnotesize
\textit{Notes:} Panel A reports the annual mean and median disruption frequency and the
share of positive firm-years over fiscal years 2006--2024. Panel B compares industries
above and below the median of the 48
industry-level employment-to-assets ratios.
\end{minipage}
\end{figure}

Figure~\ref{fig:known_groups} describes the time series and compares industries with
different reliance on labor. Disruption frequency is elevated in 2009 and again from
2020 through 2023, with the highest annual mean in 2022. Because the transcript
archives and covered firms change over time, I treat this aggregate pattern as
descriptive. The cross-sectional comparison is sharper: the mean score is 1.78
excerpts per 10,000 words in labor-intensive industries and 0.92 in other industries.

For the industry comparison, I compute the median employment-to-assets ratio within
each Fama--French 48 industry over 2006--2024. The 24 industries above the median
include restaurants and hotels, retail, personal services, healthcare, apparel,
textiles, automobiles, and construction materials; Appendix
Table~\ref{tab:app_labor_industries} reports the complete list.

The return evidence follows the same cross-sectional pattern. The interaction between
standardized disruption and the high-labor-industry indicator is 0.0085
(\(t=2.88\)) in the annual idiosyncratic-volatility regression, after controlling for
the interaction with continuous firm-level labor intensity. The result links the
strength of the risk relation to firms' reliance on the underlying production input.

\subsection{Economic stability across classification rules}

The broad construct admits reasonable disagreement at its semantic boundary. I
therefore reconstruct the score under seven rules while holding transcript sources,
denominators, fiscal alignment, samples, controls, and regression specifications
fixed. The inclusive threshold and the rule requiring agreement of at least two
classifiers accept 72.5\% and 61.4\% more excerpts than the baseline. A
higher-confidence threshold and unanimous classifier agreement accept 33.0\% and
36.9\% fewer excerpts. Despite these large differences, firm-year Spearman
correlations with the baseline range from 0.889 to 0.914. The two content exclusions
that remove shortage or leadership language have correlations above 0.97.

\begin{figure}[!htbp]
\centering
\includegraphics[width=0.92\linewidth]{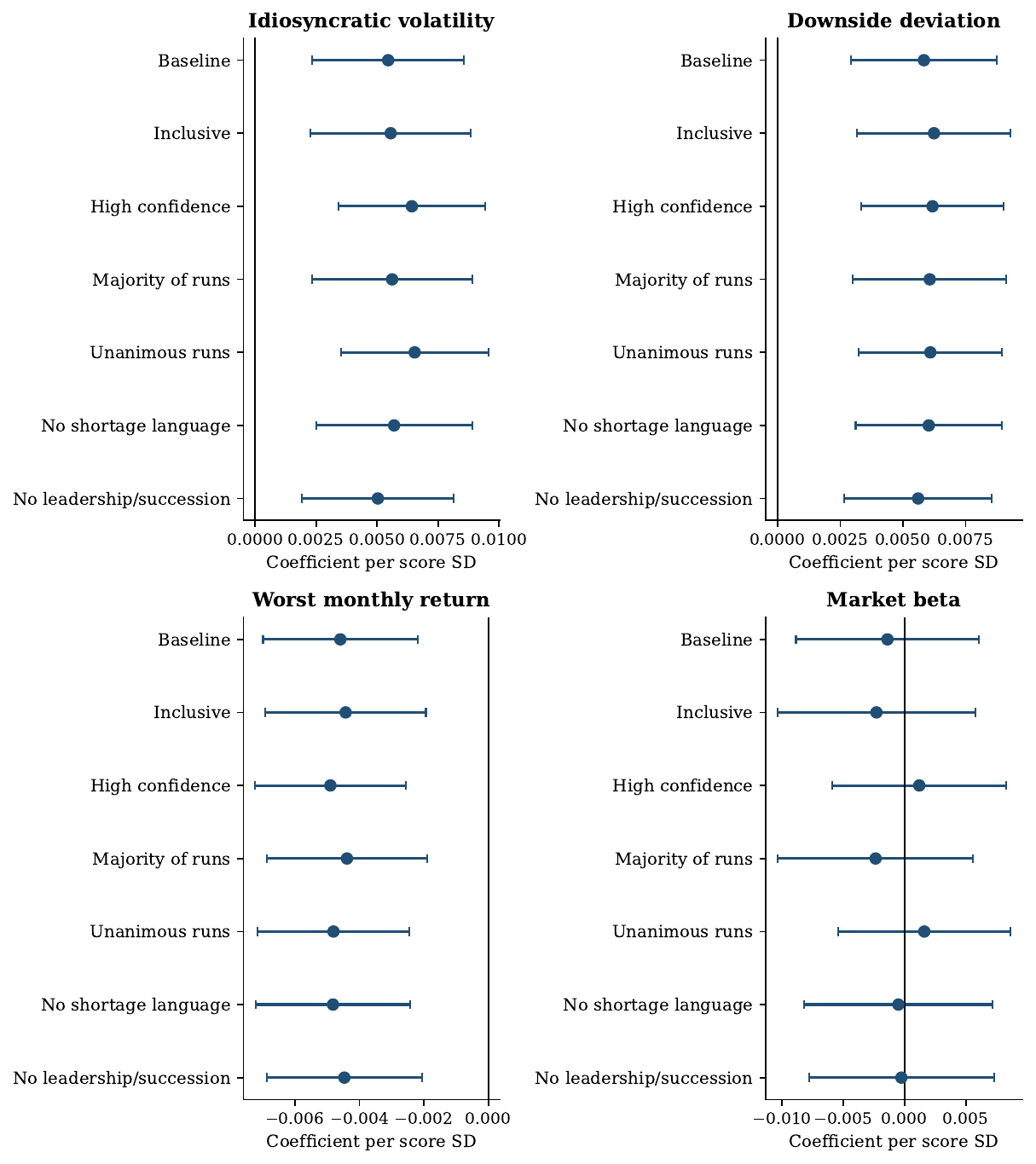}
\caption{Annual Coefficient Stability Across Classification Rules}
\label{fig:ambiguity_stability}
\begin{minipage}{0.92\linewidth}
\footnotesize
\textit{Notes:} The figure reports annual coefficients and 95\% confidence intervals
for seven broader and narrower classifications. Each specification includes broad
workforce discussion, firm controls, and firm and fiscal-year fixed effects. Samples
are held fixed across definitions within each outcome.
\end{minipage}
\end{figure}

Figure~\ref{fig:ambiguity_stability} reports the principal economic coefficients under
each rule. Idiosyncratic-volatility coefficients range from 0.00503 to 0.00654, and the
smallest absolute \(t\)-statistic is 3.16. Downside-deviation coefficients range from
0.00561 to 0.00624, with minimum \(|t|=3.73\). Worst-month coefficients range from
\(-0.00490\) to \(-0.00439\), with minimum \(|t|=3.46\). Market beta remains
statistically indistinguishable from zero under every rule. Appendix
Table~\ref{tab:app_variant_size} reports score size and rank stability, and Appendix
Table~\ref{tab:app_economic_grid} reports the complete coefficient grid.

The alternative rules materially change which excerpts are accepted, but they do not
change the ordering of the principal economic results. The firm-year evidence is thus
stable to the same boundary choices that generate disagreement on individual
excerpts.

\subsection{Construct differentiation and organizational outcomes}

The released \textcite{harford2026labor} measure identifies an important component of
the broader construct. In 25,307 common firm-calendar-quarters containing exactly one
call in this paper's sample, 62.2\% of HHQ-positive observations are also positive for
human-capital disruption, while 33.8\% of disruption-positive observations are
HHQ-positive. The continuous correlation is 0.476. Removing explicit shortage
passages reduces the correlation to 0.406. Appendix
Table~\ref{tab:hhq_overlap} reports the quarterly and annual overlap measures.

The economic comparison provides a sharper distinction. As reported in
Table~\ref{tab:hhq_incremental}, the annual disruption coefficient remains positive
after the HHQ measure and transcript-wide negative and uncertainty frequencies enter
jointly. The coefficient also remains positive after every explicit shortage passage
is removed. Market beta remains unrelated to human-capital disruption in the joint
specification, whereas the HHQ coefficient is positive. The broader score therefore
contains firm-specific risk information outside measured labor-shortage exposure and
generic adverse tone.

Outcomes measured independently of the text provide additional convergent evidence.
Table~\ref{tab:organizational} shows that, after leadership and succession passages are
removed, the measure predicts next-year named-executive roster exits and the
incumbent CEO's effective departure from office before the next call. These results
connect disclosed disruption to subsequent organizational adjustment without using
direct transition language to construct the score.

\subsection{Author guidance and classification evidence}

The measurement procedure begins with human guidance. I classified 50
workforce-related excerpts before model training and used those decisions to define
the inclusion and exclusion criteria. The final classification rule agrees with 76\%
of these examples, with Cohen's \(\kappa=0.519\) and precision and recall of 0.75.
Because the examples established the coding criteria, this exercise measures fidelity
to the author-defined boundary rather than out-of-sample performance.

A stratified audit of 100 ticker--calendar-years evaluates aggregation relative to an
Opus review of the underlying excerpts. After weighting the classifier-positive and
classifier-negative strata to their population shares, the binary any-disruption
indicator has estimated precision of 1.00 (95\% interval 0.93--1.00) and recall of
0.92 (0.87--0.96). The audit is conditional on the workforce candidate screen and
uses ticker--calendar-years, rather than the continuous gvkey--fiscal-year score in
the regressions. Appendix Tables~\ref{tab:app_measure_samples} and
\ref{tab:app_operating} report the roles of each labeled sample and the sentence-level
classifier statistics.

I also classified 120 workforce-related excerpts that Opus had reviewed. The sample
concentrates on the difficult boundary between material disruption and general
workforce discussion. The classifications agree in 65.0\% of cases, producing
Cohen's \(\kappa=0.300\), with a 95\% interval from 0.134 to 0.464
\parencite{cohen1960coefficient}. My classifications more often required an active,
unresolved adverse condition and sometimes used context beyond the three-sentence
window, whereas the written criteria also include consequential adjustments,
mitigated disruptions, and succession events.

Appendix Tables~\ref{tab:app_confusion} and \ref{tab:app_protocol} show where the
readings differ and how agreement changes under alternative applications of the
written broad-disruption rule. Because both original classifications were visible,
this exercise is a boundary-sensitivity diagnostic rather than an independent
reliability estimate. More importantly for the firm-level measure, the annual
rankings and return-distribution results remain stable under the broader and narrower
classifications reported above.

\section{Baseline Economic Evidence}
\label{sec:baseline}

\subsection{Annual specification}

The baseline annual regressions ask whether within-firm changes in disclosed
human-capital disruption coincide with changes in the distribution of the firm's
returns. For outcome \(Y_{i,t}\), the specification is
\begin{equation}
\label{eq:annual}
Y_{i,t}
=
\beta\,\widetilde{HCDisruption}_{i,t}
+\gamma\,\widetilde{HCExposure}_{i,t}
+\theta'X_{i,t}
+\alpha_i+\delta_t+\varepsilon_{i,t},
\end{equation}
where tildes denote standardization within the outcome-specific estimation sample,
\(\alpha_i\) are firm fixed effects, and \(\delta_t\) are fiscal-year fixed effects.
The controls are log assets, Tobin's q, ROA, book leverage, and cash/assets. Standard
errors are clustered by firm. The specification measures a contemporaneous
within-firm relation because call text does not predate every daily return in the
fiscal window. The annual estimates therefore describe contemporaneous within-firm
covariation; the call-timed tests below assess predictive content.

\subsection{Firm-specific return distributions}

Table~\ref{tab:annual_risk} reports the principal outcomes. A one-standard-deviation
increase in disruption is associated with 0.00545 higher annualized idiosyncratic
volatility, with \(t=3.43\). Because returns are recorded in decimals, the magnitude is
approximately 0.55 percentage points. The downside-deviation coefficient is 0.00584
(\(t=3.92\)), or 0.58 percentage points. The worst monthly return coefficient is
\(-0.00459\) (\(t=-3.75\)), implying a 0.46 percentage point more negative worst
month. These three outcomes consistently locate the relation in the firm-specific and
adverse portions of the return distribution.

\begin{table}[!htbp]
\centering
\caption{Human-Capital Disruption and the Distribution of Firm Returns}
\label{tab:annual_risk}
\small
\begin{threeparttable}
\begin{tabular}{lrrrr}
\toprule
Outcome & Coefficient & $t$-statistic & Observations & Firms\\
\midrule
Idiosyncratic volatility & 0.0055*** & 3.43 & 12,114 & 2,315\\
Downside deviation & 0.0058*** & 3.92 & 11,851 & 2,209\\
Worst monthly return & -0.0046*** & -3.75 & 11,851 & 2,209\\
Market beta & -0.0014 & -0.37 & 12,114 & 2,315\\
Fama--French three-factor explanatory share & -0.0041** & -2.22 & 12,114 & 2,315\\
\midrule
Broad workforce-discussion frequency & Yes & & &\\
Firm and fiscal-year fixed effects & Yes & & &\\
Firm-clustered standard errors & Yes & & &\\
\bottomrule
\end{tabular}
\begin{tablenotes}[flushleft]\footnotesize
\item Notes: For each regression, the disruption measure is standardized using the observations included in that regression. All models include broad workforce-discussion frequency, log assets, Tobin's q, ROA, book leverage, and cash/assets. Return-distribution outcomes and the Fama--French three-factor explanatory share are in decimal units. Therefore, the idiosyncratic-volatility coefficient of 0.00545 corresponds to approximately 0.55 percentage points per measure standard deviation. Stars denote significance at the 10\%, 5\%, and 1\% levels.
\end{tablenotes}
\end{threeparttable}
\end{table}

Market beta provides a comparison with systematic risk. Its coefficient is \(-0.00140\)
with \(t=-0.37\). The share of daily excess-return variation explained by the three
Fama--French factors is also lower, by 0.00411 with \(t=-2.22\), while total daily
volatility increases. The combination of higher idiosyncratic risk and an
insignificant market-beta coefficient is consistent with firm-specific rather than
systematic risk.

The annualized idiosyncratic-volatility mean is 0.318 in the descriptive sample, so
the 0.00545 coefficient equals approximately 1.7\% of the outcome mean. The magnitude
represents a within-firm shift associated with a disclosed organizational condition.

\subsection{Specification and sample sensitivity}

Two-way clustering by firm and year reduces the idiosyncratic-volatility \(t\)-statistic
from 3.43 to 2.86 without changing the coefficient. Excluding 2020--2021 yields a
coefficient of 0.0053 with \(t=2.68\). Estimates are positive in 2006--2014 and
2015--2024 separately, with \(t=2.46\) and 1.95. Replacing the raw standardized
frequency with an asinh transformation gives \(t=3.59\), and winsorizing the outcome
at 0.5 and 99.5 percent gives \(t=3.42\). The maximum-F1 score, the disruption share
among workforce candidates, and the positive-intensity subsample also yield positive
idiosyncratic-volatility coefficients. Appendix Table~\ref{tab:app_annual_robustness}
reports these estimates.

The extensive-margin indicator is weaker in the annual model (\(t=1.12\)). This
contrasts with the CEO-office-exit analysis, where an any-positive call indicator is
more informative than intensity conditional on a positive score. I therefore use the
continuous frequency in the annual return regressions and report both functional forms
for the rare call-level exit outcome.

A cross-sectional specification with Fama--French 48 industry and year fixed effects,
rather than firm fixed effects, yields an idiosyncratic-volatility coefficient close to
zero. The annual result is concentrated in changes within a firm over time rather than
in a stable cross-sectional ranking of firms.

\subsection{Labor intensity}

The labor-intensity interaction provides cross-sectional evidence. The coefficient on
disruption interacted with the high-labor-industry indicator is 0.0085
(\(t=2.88\)) in the annual idiosyncratic-volatility model. The result is consistent
with the same disruption carrying more economic weight when labor is a more important
production input. Because industry labor intensity also reflects production
technology and operating structure, I use the interaction as construct-validity
evidence rather than as a separate source of identification.

\subsection{Incremental content beyond labor shortages and generic tone}

Table~\ref{tab:hhq_incremental} compares the human-capital measure with both the
released \textcite{harford2026labor} measure and standard transcript-wide
Loughran--McDonald negative and uncertainty frequencies
\parencite{loughran2011liability}. The tone variables are measured from the exact
transcripts in this paper and aggregate the full call rather than only
workforce-related passages.

\begin{table}[!htbp]
\centering
\caption{Human-Capital Disruption and Labor-Shortage Exposure in Joint Risk Models}
\label{tab:hhq_incremental}
\footnotesize
\begin{threeparttable}
\begin{tabularx}{\linewidth}{Xcccr}
\toprule
Measure and controls & Tone & HC coefficient & HHQ coefficient & $N$\\
\midrule
\multicolumn{5}{l}{\textit{Panel A. Annualized idiosyncratic volatility}}\\
HC measure alone & No & 0.0081*** (4.02) & -- & 7,019\\
HHQ measure alone & No & -- & -0.0014 (-0.74) & 7,019\\
HC and HHQ & No & 0.0096*** (4.25) & -0.0049** (-2.20) & 7,019\\
HC and HHQ with generic tone & Yes & 0.0065*** (2.97) & -0.0047** (-2.16) & 7,019\\
HC excluding shortage passages, HHQ, and tone & Yes & 0.0068*** (3.28) & -0.0044** (-2.14) & 7,019\\
\midrule
\multicolumn{5}{l}{\textit{Panel B. Annual market beta}}\\
HC and HHQ with generic tone & Yes & 0.0062 (1.30) & 0.0124** (2.30) & 7,019\\
\midrule
\multicolumn{5}{l}{\textit{Panel C. Log idiosyncratic volatility over the next 42 trading days}}\\
All calls, HC measure alone & No & 0.0049** (2.09) & -- & 41,009\\
All calls, HC measure with generic tone & Yes & 0.0033 (1.59) & -- & 41,009\\
One-call HHQ sample, HC, HHQ, and tone & Yes & 0.0065** (2.17) & -0.0061** (-2.24) & 23,673\\
One-call HHQ sample, no shortage language & Yes & 0.0055* (1.80) & -0.0053** (-1.97) & 23,673\\
\bottomrule
\end{tabularx}
\begin{tablenotes}[flushleft]\footnotesize
\item Notes: HHQ denotes the Harford--He--Qiu labor-shortage exposure measure. Coefficients are reported per sample standard deviation, with \(t\)-statistics in parentheses. Panel A uses 7,019 common firm-year observations from 2006--2021 and includes firm and year fixed effects, broad workforce discussion, and firm controls. Its dependent variable is annualized idiosyncratic volatility in decimal units. Panel B uses market beta in levels. Joint HC and HHQ coefficients are conditional decompositions of correlated text measures. The annual comparison aligns HHQ calendar-year observations with this paper's fiscal-year observations; this retains a within-year timing mismatch. Panel C uses log idiosyncratic volatility over the next 42 trading days; multiplying coefficients by 100 gives the approximate percentage change. It adds pre-call risk and outcome-quarter fixed effects. The one-call HHQ sample contains common firm-calendar-quarters with exactly one call in this paper's transcript sample. Because the HHQ data omit call dates and transcript text, this is a quarter-level comparison rather than a verified same-transcript match. Generic tone controls are transcript-wide Loughran--McDonald negative and uncertainty word frequencies. Annual standard errors are clustered by firm; Panel C standard errors are clustered by firm and outcome quarter. Stars denote significance at the 10\%, 5\%, and 1\% levels.
\end{tablenotes}
\end{threeparttable}
\end{table}

In the common annual sample, the disruption coefficient is 0.00806
(\(t=4.02\)) without either comparison measure and 0.00962 (\(t=4.25\)) after adding
HHQ. Adding negative and uncertainty frequencies reduces the coefficient to 0.00651
(\(t=2.97\)), or approximately 0.65 percentage point of annualized idiosyncratic
volatility. The negative-tone coefficient is 0.02241 (\(t=7.36\)), indicating that
general adverse language is itself associated with firm-specific risk. Nevertheless, deleting
explicit shortage passages leaves a disruption coefficient of 0.00683
(\(t=3.28\)) in the fully controlled model.

The HHQ idiosyncratic-volatility coefficient is small when entered alone
(\(-0.00141\), \(t=-0.74\)) and negative in the joint model with tone
(\(-0.00467\), \(t=-2.16\)). I treat the joint signs as a decomposition of correlated
text measures. The relevant comparison is whether the broader disruption score
retains incremental firm-specific risk information, which it does.

The risk distinction also appears in market beta. In the specification with HHQ and
generic tone, the disruption coefficient is 0.00615 (\(t=1.30\)), while the HHQ
coefficient is 0.01236 (\(t=2.30\)). Together with the idiosyncratic-volatility
results, this contrast shows that the two text measures are not interchangeable in
their empirical relation to systematic and firm-specific risk.

The call-level comparison is less precise. Across all transcripts in this paper,
adding generic tone reduces the 42-day idiosyncratic-volatility coefficient from
0.00491 (\(t=2.09\)) to 0.00332 (\(t=1.59\)). Generic tone therefore explains part of
the short-horizon relation. In common firm-quarters containing exactly one call, the
coefficient is 0.00653 (\(t=2.17\)) after both HHQ and tone are included, and 0.00553
(\(t=1.80\)) after shortage passages are also removed. The annual estimates provide
the strongest evidence of incremental firm-specific risk information, while the
call-level comparison shows partial overlap with generic adverse language.

\subsection{Persistence and longer-horizon outcomes}

The score itself is persistent. In a firm-fixed-effect autoregression, the lagged
standardized score has a coefficient of 0.148 with \(t=4.64\). This persistence
motivates the call-level timing analysis.

Broad next-year outcomes are weaker. Appendix Table~\ref{tab:app_long_horizon} reports
statistically imprecise coefficients for next-year idiosyncratic and total volatility,
employment growth, sales growth, capital expenditure, ROA, and 12-month returns.
Absolute employment growth is also unrelated to the score, while the coefficient on
absolute sales growth is negative and marginally significant. The current annual
specifications therefore do not detect a common directional change in average real
outcomes.

The annual evidence is strongest for the distribution of firm-specific returns. The
next section uses call timing to test whether the score also contains information
about subsequently realized risk.

\section{Dynamic, Predictive, and Organizational Evidence}
\label{sec:dynamic}

The call-level analysis shows that human-capital disruption also contains information
about subsequently realized firm-specific risk. I first measure risk in matched
trading-day windows before and after each earnings call. I then examine executive
turnover and CEO-office exit as organizational outcomes measured after the disclosure
period. The timing tests distinguish near-term prediction from contemporaneous
association.

\subsection{Call-timed firm-specific risk}
\label{subsec:call_timed}

The call-level design asks whether the current call contains information about the
subsequent level of firm-specific risk beyond the same firm's matched pre-call
realization. The unit of observation is an earnings call, unique by firm and
call date. I define the event day as the first CRSP trading day on or after the call
date. The pre-call window ends two trading days before the event day, and the
post-call window begins two trading days afterward, thereby excluding the immediate
call-response interval. Daily idiosyncratic-risk outcomes are constructed from
residuals of the Fama--French three-factor model \parencite{fama1993common}.

For positive risk outcomes, the primary specification is
\begin{equation}
\label{eq:call}
\begin{aligned}
\log\!\left(Risk^{\,post,h}_{i,c}\right)
={}&
\beta_h\,\widetilde{HCDisruption}_{i,c}
+\rho_h\log\!\left(Risk^{\,pre,h}_{i,c}\right)
+\gamma_h\,\widetilde{HCExposure}_{i,c}\\
&+X_{i,c^-}'\theta_h
+\alpha_i+\lambda_{q(c,h)}+\varepsilon_{i,c,h},
\end{aligned}
\end{equation}
where \(i\) indexes firms, \(c\) indexes calls, and \(h\) denotes the post-call
horizon. The disruption and broad-exposure variables are frequencies per 10,000
transcript words and are standardized within the estimation sample. The vector
\(X_{i,c^-}\) contains the latest annual accounting information released strictly
before the call. The specification includes firm fixed effects and outcome-window
calendar-quarter fixed effects. Standard errors are two-way clustered by firm and
outcome-window quarter, following the general multiway-clustering logic of
\textcite{cameron2011robust}.

Panel A of Table~\ref{tab:call_timed} reports two nonoverlapping 21-trading-day
blocks. Conditional on the nearest pre-call block, a one-standard-deviation increase
in disruption disclosure is associated with 0.54\% higher idiosyncratic volatility in
trading days \(+2\) through \(+22\), with \(t=2.14\), and 0.56\% higher
idiosyncratic volatility in trading days \(+23\) through \(+43\), with \(t=2.09\).
The downside-deviation coefficients are 0.36\% in the first block
(\(t=1.38\)) and 0.62\% in the second (\(t=2.30\)). Thus, the
idiosyncratic-volatility result appears in both of the first two blocks, whereas the
downside result emerges only in the second.

The cumulative 42-trading-day specification gives a similar result. A
one-standard-deviation increase in disruption predicts 0.49\% higher idiosyncratic
volatility in the full sample (\(t=2.18\)) and 0.50\% (\(t=2.05\)) when the outcome
window is required to end before the next observed call. Downside estimates are
positive but less precise, tail-loss estimates are imprecise, and market beta remains
unrelated to the score. I therefore treat 42-day idiosyncratic volatility as the
principal call-level result; secondary outcomes and content exclusions appear in the
Appendix.

\begin{table}[!htbp]
\centering
\caption{Human-Capital Disruption and Subsequent Firm-Specific Risk}
\label{tab:call_timed}
\footnotesize
\begin{threeparttable}
\begin{tabularx}{\linewidth}{Xlrrrr}
\toprule
Horizon and sample & Outcome & Association per SD & $t$ & $N$ & Firms\\
\midrule
\multicolumn{6}{l}{\textit{Panel A. Nonoverlapping 21-trading-day blocks}}\\
Trading days +2 to +22 & Idiosyncratic volatility & 0.54\%** & 2.14 & 41,559 & 2,510\\
Trading days +2 to +22 & Downside deviation & 0.36\% & 1.38 & 41,559 & 2,510\\
Trading days +23 to +43 & Idiosyncratic volatility & 0.56\%** & 2.09 & 41,559 & 2,510\\
Trading days +23 to +43 & Downside deviation & 0.62\%** & 2.30 & 41,559 & 2,510\\
\midrule
\multicolumn{6}{l}{\textit{Panel B. Cumulative 42-trading-day window}}\\
All calls & Idiosyncratic volatility & 0.49\%** & 2.18 & 41,787 & 2,527\\
All calls & Downside deviation & 0.42\%* & 1.86 & 41,787 & 2,527\\
Window ends before the next observed call & Idiosyncratic volatility & 0.50\%** & 2.05 & 38,357 & 2,389\\
Window ends before the next observed call & Downside deviation & 0.44\%* & 1.79 & 38,357 & 2,389\\
\bottomrule
\end{tabularx}
\begin{tablenotes}[flushleft]\footnotesize
\item Notes: Post-call risk windows begin at trading day +2, omitting the immediate call-response interval. Models include matched pre-call risk, a control for broad workforce discussion, strictly pre-call accounting controls, firm fixed effects, and outcome-quarter fixed effects; standard errors are two-way clustered by firm and calendar quarter. The outcomes are logged, so the reported associations are coefficient-implied percentage differences per standard deviation of the disruption score. The restricted 42-day sample requires an observed next call and retains only windows ending strictly before that call; 97.9\% of valid 42-day windows with an observed successor call satisfy this restriction. Appendix Table~\ref{tab:app_call_secondary} reports secondary outcomes, content exclusions, and 63-day estimates.
\end{tablenotes}
\end{threeparttable}
\end{table}

I use log risk because the positive risk measures are strongly right-skewed.

At 63 trading days, idiosyncratic volatility is 0.69\% higher per measure standard
deviation (\(t=3.22\)). Because only 30.3\% of these windows end before the next
observed call, I interpret the 63-day estimates as evidence of persistence across
reporting periods and retain 42 days as the primary predictive horizon. Complete
63-day estimates appear in Appendix Table~\ref{tab:app_call_secondary}.

\subsection{The dynamic risk path}
\label{subsec:dynamic_path}

\begin{figure}[!htbp]
\centering
\includegraphics[width=0.94\linewidth]{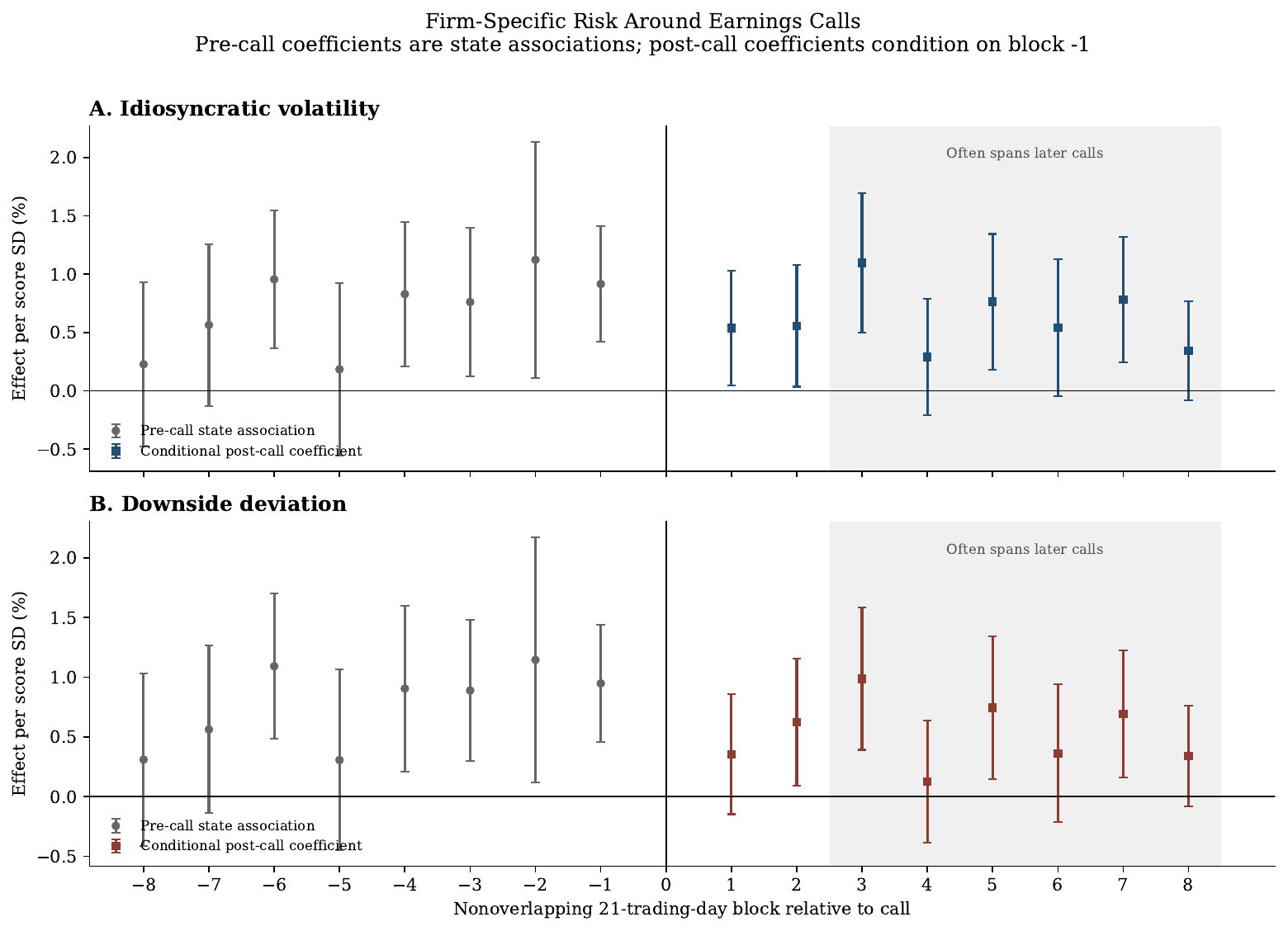}
\caption{Firm-Specific Risk Around Earnings Calls}
\label{fig:dynamic_risk}
\begin{minipage}{0.94\linewidth}
\footnotesize
\textit{Notes:} The figure reports percentage effects per standard deviation of the
call-level disruption score. Pre-call blocks are state associations from models with
firm and block-outcome-quarter fixed effects. Post-call blocks condition on the
nearest pre-call risk block, broad workforce exposure, and strictly pre-call public
controls. Because the pre- and post-call points come from different specifications,
they are shown as separate marker series and are not connected. Confidence intervals
use firm and block-outcome-quarter clustered standard errors. The shaded later region
increasingly spans subsequent calls.
\end{minipage}
\end{figure}

Figure~\ref{fig:dynamic_risk} plots nonoverlapping 21-trading-day blocks before and
after each call on a common sample of 41,559 calls and 2,510 firms. Trading-day
positions \(-1\), \(0\), and \(+1\) are omitted.

The risk state is visible well before the call. Grouping the eight pre-call blocks
into far, middle, and near periods yields idiosyncratic-volatility coefficients of
0.63\% (\(t=2.25\)), 0.69\% (\(t=2.49\)), and 0.85\% (\(t=3.36\)),
respectively. The corresponding downside-deviation coefficients are 0.71\%
(\(t=2.47\)), 0.80\% (\(t=2.76\)), and 0.87\% (\(t=3.44\)). There is no
visible discontinuity at the call date.

Post-minus-pre changes are small: the \(t\)-statistics are 0.59 for idiosyncratic
volatility, \(-1.03\) for downside deviation, and \(-0.36\) for tail-loss magnitude.
These estimates complement the conditional post-call regressions. The call contains
information about the subsequent risk level after controlling for pre-call risk, but
the risk state does not first emerge at disclosure. Together, the positive pre-call
path, the first two post-call blocks, and the null changes characterize a persistent
organizational-risk state.

\subsection{Subsequent organizational outcomes}
\label{subsec:organizational}

I next examine whether the measure has predictive content for organizational
adjustment. Employee mobility and key-person departures can destroy firm-specific
knowledge \parencite{campbell2012who,israelsen2017key}, while CEO turnover itself is
related to performance and managerial fit
\parencite{jenter2015ceo,jenter2021performance}. These outcomes provide evidence from
inside the organization that differs from the return-based variables used above.

\paragraph{Annual executive-roster turnover.}
I rebuild the firm-year score after removing every accepted passage that mentions
leadership or succession. I then estimate firm- and year-fixed-effect models for
outcomes measured in the exact next fiscal year, with broad workforce discussion and
standard firm controls.

Panel A of Table~\ref{tab:organizational} shows that the no-leadership score predicts a
0.96-percentage-point higher named-executive roster-exit rate (\(t=3.17\)), a
2.79-percentage-point higher probability of any roster exit (\(t=3.34\)), and a
1.13-percentage-point higher probability of CEO roster absence (\(t=2.79\)). The CEO
coefficient is roughly half the estimate obtained from the unrestricted score, while
all three no-leadership estimates remain positive. Roster exit denotes absence from
the next named-executive roster; the exact CEO-office-exit test follows separately.

\paragraph{CEO-office exit before the next call.}
Panel B of Table~\ref{tab:organizational} examines whether the incumbent CEO leaves
office strictly before the firm's next observed earnings call. The risk set uses
nonoverlapping call-to-next-call intervals of 45 to 150 days, requires exactly one CEO
spell active on the focal call date, treats missing future coverage as censoring, and
uses the no-leadership disruption score. The fully controlled sample contains 27,691
intervals, 715 exits, and 1,411 firms.

A one-standard-deviation increase in the no-leadership score predicts a
0.391-percentage-point higher probability of CEO-office exit before the next call
(\(t=2.53\)), approximately 15\% of the 2.54\% event rate.
Complementary-log-log and conditional-Poisson specifications produce ratios of 1.088
and 1.112, respectively. An any-positive indicator predicts a
0.949-percentage-point higher probability (\(t=3.74\)), placing most of the
association on the extensive margin.

\begin{landscape}
\begin{table}[!htbp]
\centering
\caption{Human-Capital Disruption and Subsequent Executive Turnover}
\label{tab:organizational}
\fontsize{7.2}{8.0}\selectfont
\renewcommand{\arraystretch}{0.88}
\setlength{\tabcolsep}{4pt}
\begin{threeparttable}
\begin{tabular}{p{3.8cm}p{5.0cm}rrrrrr}
\toprule
Outcome or specification & Human-capital measure & Estimate & $t$ or 95\% CI & $N$ & Firms & Events & Intervals omitted\\
\midrule
\multicolumn{8}{l}{\textit{Panel A. Next-year Execucomp roster outcomes}}\\
CEO roster absence & Human-capital disruption & 2.091 pp*** & 4.25 & 9,962 & 1,458 & & \\
CEO roster absence & Human-capital disruption excluding leadership/succession passages & 1.129 pp*** & 2.79 & 9,962 & 1,458 & & \\
Named-executive roster-exit rate & Human-capital disruption & 1.510 pp*** & 4.54 & 10,000 & 1,466 & & \\
Named-executive roster-exit rate & Human-capital disruption excluding leadership/succession passages & 0.962 pp*** & 3.17 & 10,000 & 1,466 & & \\
Any named-executive roster exit & Human-capital disruption & 4.090 pp*** & 4.71 & 10,000 & 1,466 & & \\
Any named-executive roster exit & Human-capital disruption excluding leadership/succession passages & 2.786 pp*** & 3.34 & 10,000 & 1,466 & & \\
\midrule
\multicolumn{8}{l}{\textit{Panel B. Effective CEO-office exit before the next call}}\\
Linear probability model, full sample & Human-capital disruption excluding leadership/succession passages, per SD & 0.391 pp** & 2.53 & 27,691 & 1,411 & 715 & 0\\
Complementary-log-log, full sample & Human-capital disruption excluding leadership/succession passages, per SD & 1.088*** & [1.029, 1.151] & 27,691 & 1,411 & 715 & 0\\
Conditional Poisson, firm FE & Human-capital disruption excluding leadership/succession passages, per SD & 1.112** & [1.009, 1.225] & 19,418 & 480 & 715 & 0\\
Linear probability model, exclude specific prior disclosure & Human-capital disruption excluding leadership/succession passages, per SD & 0.481 pp*** & 3.82 & 27,506 & 1,411 & 530 & 185\\
Complementary-log-log, exclude specific prior disclosure & Human-capital disruption excluding leadership/succession passages, per SD & 1.127*** & [1.067, 1.190] & 27,506 & 1,411 & 530 & 185\\
Linear probability model, exclude broad prior/same-day disclosure & Human-capital disruption excluding leadership/succession passages, per SD & 0.237 pp** & 2.50 & 27,290 & 1,411 & 314 & 401\\
Complementary-log-log, exclude broad prior/same-day disclosure & Human-capital disruption excluding leadership/succession passages, per SD & 1.145*** & [1.065, 1.231] & 27,290 & 1,411 & 314 & 401\\
\bottomrule
\end{tabular}
\begin{tablenotes}[flushleft]\scriptsize
\item Notes: Panel A coefficients are percentage-point differences per sample standard deviation; standard errors are two-way clustered by firm and fiscal year. ``Roster exit'' means absence from the next annual named-executive roster, not a confirmed company-departure date. Panel B defines an event as the incumbent's effective exit from the CEO office strictly before the next observed call. Linear-probability coefficients are percentage-point differences per standard deviation. Complementary-log-log models report exponentiated coefficients as discrete-time hazard ratios; the conditional Poisson model reports an incidence-rate ratio. The specific prior-disclosure restriction omits 185 intervals when a pre-call SEC filing contains directional incumbent-transition language and a nearby effective date. The broader restriction omits 401 intervals when pre-call or same-day filings contain broader incumbent-transition language. These restrictions reduce cases in which the exit may already have been disclosed, but they do not classify every possible prior disclosure. Stars denote significance at the 10\%, 5\%, and 1\% levels.
\end{tablenotes}
\end{threeparttable}
\end{table}
\end{landscape}

Timing provides additional information. The current call's measure is only
weakly related to the preceding interval's exit (\(t=1.45\)), but the next call's
measure predicts the current interval's exit (\(t=2.74\)). When current and
prior-call scores enter jointly, the prior score is statistically significant and the
current score is marginal. The text therefore appears to track a persistent
organizational condition rather than the first disclosure of an unanticipated CEO
transition.

The estimate remains positive after removing the five most influential firms but
becomes imprecise under aggressive observation-level Cook-distance trimming; Appendix
Table~\ref{tab:app_ceo_diagnostics} reports these tests.

\paragraph{Pre-call public-information screens.}
Execucomp's \texttt{LEFTOFC} variable records the effective date on which an
individual left the CEO office. It is neither an announcement date nor a
forced-versus-voluntary classification. I therefore conduct a targeted Item 5.02 audit
of 3,270 exit-document pairs, representing 3,135 distinct pre-call Item 5.02 filings
associated with the 715 exits.
The conservative screen identifies 185 exits for which a pre-call filing
contains directional evidence of the incumbent's CEO transition and a nearby date
matching the effective office-exit date.

Removing these 185 previously disclosed transitions leaves a coefficient of 0.481
percentage points (\(t=3.82\)). A broader sensitivity test excludes 401 of the 715
exits whenever pre-call or same-day filings contain a wider set of names, CEO terms,
transition language, and relevant dates. It leaves a coefficient of 0.237 percentage points
(\(t=2.50\)). The corresponding pooled complementary-log-log ratio is 1.145
([1.065, 1.231]), and the conditional-Poisson ratio is 1.171 ([1.026, 1.336]).

Together, the no-leadership score and filing screens show that the organizational
association extends beyond direct succession language and readily identified prior
Item 5.02 disclosures. The analysis does not classify departures as forced,
voluntary, or previously unannounced; other SEC forms, press releases, news reports,
and earlier disclosures remain outside the filing screen.

\section{Robustness and Scope}
\label{sec:robustness}

\subsection{Alternative construction, samples, and inference}

The annual idiosyncratic-volatility result remains positive under alternative
confidence thresholds, classifier-agreement rules, transformations, winsorization,
and score trimming. Two-way clustering by firm and year produces \(t=2.86\);
excluding 2020--2021 produces \(t=2.68\); and the estimates are positive in the
2006--2014 and 2015--2024 subsamples. Section~\ref{sec:validation} and Appendix
Table~\ref{tab:app_annual_robustness} report the coefficient comparisons.

At the call level, the 42-day idiosyncratic-volatility coefficient remains positive
after excluding 2020--2021 and after replacing calendar-quarter fixed effects with
Fama--French industry-by-quarter fixed effects.

\subsection{Workforce salience, labor shortages, and adverse tone}

Every principal specification controls for broad workforce discussion. In the annual
sample, the disruption coefficient remains positive after adding the HHQ
labor-shortage measure and transcript-wide negative and uncertainty frequencies and
after deleting all explicit shortage passages. At the call level, generic tone reduces
the 42-day coefficient from 0.00491 (\(t=2.09\)) to 0.00332 (\(t=1.59\)); in the
common one-call-quarter sample, the coefficient is 0.00653 (\(t=2.17\)) with HHQ and
tone included. The annual comparison therefore provides the stronger evidence of
incremental content.

Section~\ref{subsec:organizational} reports the corresponding content exclusion for
executive outcomes: the no-leadership score continues to predict named-executive
roster exit and CEO-office exit.

\subsection{Measurement and sample boundaries}

The score captures disclosed disruptions expressed through the workforce-language
screen. Among vocabulary-matched sentences rejected by the preliminary classifier, a
stratified audit implies a population-weighted miss rate of 2.15\% and estimated
screen recall of approximately 0.73. The regression standard errors condition on the
estimated score; the seven-rule exercise evaluates plausible classification boundaries
but does not resample the complete measurement procedure.

Coverage also requires an observed earnings call, valid accounting and security links,
and sufficient return data. The combined transcript archives place greater weight on
large public firms and vary in coverage over time. The HHQ comparison additionally
matches a calendar-year measure to this paper's fiscal-year measure, with exact
alignment only for December-year-end firms. These features define the population to
which the estimates apply.

Finally, firms choose what to discuss, and both the text and the outcomes may respond
to omitted operating conditions. Firm fixed effects, calendar-time controls, matched
pre-call risk, content exclusions, and public-information screens sharpen the economic
interpretation but do not create exogenous variation in human-capital disruption.

\section{Conclusion}
\label{sec:conclusion}

This paper constructs a firm-level measure of disclosed human-capital disruption and
documents that it tracks an economically important firm-specific risk state. The measure
captures material disturbances to workforce availability, cost, skills, safety,
continuity, and organization. It is broader than labor-shortage exposure and narrower
than general workforce discussion.

Within firms, periods with greater disclosed disruption have higher idiosyncratic
volatility and downside deviation and a more negative worst monthly return. Market
beta is unrelated to the measure. The return-distribution evidence is stable across
classification rules that accept 37\% fewer to 72\% more excerpts, and the result
remains after explicit shortage passages are removed and the released
\textcite{harford2026labor} measure and generic adverse tone are included. The
stronger relation in labor-intensive industries further connects the score to firms'
reliance on human capital as a production input.

The call-timed and organizational evidence extends the result beyond contemporaneous
annual covariation. Disruption predicts firm-specific risk over each of the next two
21-trading-day blocks after conditioning on pre-call risk. The risk state is also
visible before the call, indicating persistence rather than a disclosure-date jump.
After leadership and succession passages are removed, the measure predicts subsequent
named-executive roster exits and the incumbent CEO's effective departure from office.

Together, the results show that earnings calls reveal disturbances to a key
organizational input that are distinct from general employee discussion, extend
beyond labor shortages, and contain information about the distribution and persistence
of firm-specific risk.

\clearpage
\printbibliography

\clearpage
\appendix
\section{Construct Scope and Coding Rules}
\label{app:construct}

This appendix records the operational boundary used throughout the paper. It is
included to make clear that ``human-capital disruption'' is neither a synonym for
employee discussion nor a signed measure of managerial sentiment. The classification
unit is a workforce-vocabulary candidate sentence together with one sentence on
either side. The accepted event remains attributed to the candidate sentence so that
overlapping windows are not counted multiple times merely because they share context.

\begin{table}[H]
\centering
\caption{Operational Construct Rubric}
\label{tab:app_rubric}
\footnotesize
\begin{threeparttable}
\begin{tabularx}{\linewidth}{p{3.2cm}XX}
\toprule
Family & Positive boundary & Excluded boundary\\
\midrule
Availability and skills &
Firm-specific hiring difficulty, labor shortage, absenteeism, skill scarcity, or
staffing constraint with operational or cost consequence &
Generic labor-market commentary, recruiting strategy without a constraint, and
discussion of another entity's workforce\\
Attrition and retention &
Material turnover, retention pressure, key-person dependence, or continuity risk &
Low or declining attrition, strong-retention reassurance, and customer or inventory
``turnover''\\
Cost and compensation &
Wage inflation, labor-cost repricing, furlough, pay reduction, or benefit pressure
that bears on the reporting firm &
Routine compensation metrics, immaterial changes, or statements that labor-cost
pressure is absent\\
Workforce adjustment &
Material layoff, reduction in force, hiring freeze, redeployment, restructuring, or
integration with stated operational stakes &
Routine headcount accounting, ordinary hiring progress, or generic workforce
strategy\\
Safety and continuity &
Health, safety, strike, union, or continuity conditions that impair availability or
operations &
Boilerplate safety priorities and generic employee-wellness statements\\
Leadership and succession &
Material executive succession, departure, search, or management reorganization &
Unsupported analyst probes and references that do not concern the reporting firm's
leadership\\
Forward uncertainty &
Conditional human-capital exposure that may affect costs, operations, investment, or
guidance &
Pure reassurance that a disruption did not occur or no longer exists, unless the
window still documents a consequential event or adjustment\\
\bottomrule
\end{tabularx}
\begin{tablenotes}[flushleft]\footnotesize
\item Notes: A positive case requires firm-specific material disturbance,
constraint, cost pressure, uncertainty, or consequential adjustment. Managerial tone
does not determine the label. Mitigation can remain positive when it documents the
underlying disturbance, but reassurance that no disruption existed is negative.
\end{tablenotes}
\end{threeparttable}
\end{table}

\begin{table}[!htbp]
\centering
\caption{Illustrative Transcript Excerpts}
\label{tab:app_examples}
\footnotesize
\begin{threeparttable}
\begin{tabularx}{\linewidth}{p{2.5cm}Xp{4.2cm}}
\toprule
Classification & Short excerpt & Economic reading\\
\midrule
Positive: availability &
WRK (Aug.\ 4, 2022): ``The tight labor market continues to pose challenges.'' &
Firm-specific labor availability and retention pressure\\
Positive: workforce adjustment &
Ford (July 25, 2008): ``We recorded a charge of \$274 million associated with
personnel separation programs in North America.'' &
Material workforce reduction with a quantified cost\\
Positive: resolved adjustment &
KMX (Apr.\ 12, 2022): ``We had some staffing challenges and we ramped up that
staffing throughout the year.'' &
The firm describes the underlying constraint and the costly response even though
staffing later improved\\
Negative: general discussion &
GNW (Feb.\ 3, 2012): ``[We] made a number of moves on the people front to strengthen
the team and our capabilities.'' &
Generic organizational improvement without a material disturbance\\
Negative: another entity &
AMN (Nov.\ 4, 2021): workforce-technology products ``alleviate the labor shortages.'' &
The shortages belong to customers and increase demand for the reporting firm's
products\\
Boundary: mitigated condition &
HD (Aug.\ 15, 2023): ``More consistent staffing levels'' improve service and safety
following a wage investment. &
The broad rule counts the costly action taken to stabilize staffing; the original
author and Opus classifications differed\\
Boundary: general labor market &
VMI (July 22, 2021): ``the war for talent is pervasive and competitive.'' &
The broad rule excludes the statement because the excerpt gives no firm-specific
constraint or consequence; the original raters differed\\
\bottomrule
\end{tabularx}
\begin{tablenotes}[flushleft]\footnotesize
\item Notes: Excerpts are shortened for display; classification uses the complete
candidate sentence and its adjacent context. The first five receive the same label
under conservative, central, and inclusive applications of the written rule. The
final two illustrate protocol-sensitive boundaries in the 120-excerpt author--Opus
comparison.
\end{tablenotes}
\end{threeparttable}
\end{table}

The binary boundary is intentional. Severity, tone, resolution, and content family
could each be useful additional dimensions, but each would introduce a new judgment
that requires separate reliability and economic validation. Content-family flags are
therefore used for exclusions and interpretation, not aggregated into a multidimensional
headline score.

\section{Classifier Training and Coverage of the Candidate Screen}
\label{app:model}

Table~\ref{tab:app_measure_samples} separates the samples by role. My 50
classifications informed the coding criteria but were not training observations. The
729 Opus labels trained DeBERTa, while the 400-excerpt evaluation sample selected the
classification thresholds. The 120-excerpt author--Opus comparison came later and did
not change the reported measure.

\begin{table}[!htbp]
\centering
\caption{Measurement Samples and Their Roles}
\label{tab:app_measure_samples}
\footnotesize
\begin{threeparttable}
\begin{tabularx}{\linewidth}{p{3.0cm}p{2.2cm}p{2.8cm}X}
\toprule
Sample & Size & Positive count or stratum & Purpose\\
\midrule
Author-classified examples & 50 excerpts & 24 author positives & Establish the coding
criteria and compare them with the final classifier\\
Opus training sample & 729 & 274 & Labels used to fine-tune each DeBERTa model\\
Opus evaluation sample & 400 & 219 & Ranking performance and threshold selection; exact
excerpts are disjoint from model fitting\\
Classifier application sample & 173,273 excerpts & 49,840 exceed the primary threshold &
Classification of workforce-related candidate excerpts before aggregation to calls
and firm-years\\
Annual aggregation comparison & 100 ticker--calendar-years & 50 classifier-positive;
50 classifier-negative & Population-weighted fidelity of the annual any-disruption
indicator to Opus classifications\\
Author--Opus boundary-case sample & 120 & 46 author; 60 Opus & Agreement on excerpts
identified by the broad workforce-content screen\\
\bottomrule
\end{tabularx}
\begin{tablenotes}[flushleft]\footnotesize
\item Notes: The three DeBERTa models use random seeds 0, 1, and 2. ``Positives'' refer
to the rater or threshold relevant to each row and therefore are not directly comparable
across rows.
\end{tablenotes}
\end{threeparttable}
\end{table}

\begin{table}[!htbp]
\centering
\caption{Sentence Classification Performance and Thresholds}
\label{tab:app_operating}
\footnotesize
\begin{threeparttable}
\begin{tabular}{lrrrr}
\toprule
Model or threshold & Threshold & AP & AUC & Precision / recall\\
\midrule
DeBERTa model 1 & & 0.845 & 0.832 &\\
DeBERTa model 2 & & 0.868 & 0.853 &\\
DeBERTa model 3 & & 0.849 & 0.827 &\\
Three-model ensemble & & 0.864 & 0.850 &\\
Primary high-precision rule & 0.9208 & & & 0.901 / 0.585\\
Alternative \(F_1\)-maximizing rule & 0.6492 & & & 0.813 / 0.813\\
\bottomrule
\end{tabular}
\begin{tablenotes}[flushleft]\footnotesize
\item Notes: Average precision (AP) and area under the ROC curve (AUC) are calculated
against the 400 Opus-labeled evaluation excerpts. The primary threshold was chosen as
the highest-recall threshold with precision of at least 0.90. The same sample selected
both reported thresholds.
\end{tablenotes}
\end{threeparttable}
\end{table}

The disruption classifier is applied after a broad workforce-content screen. An
excerpt enters the candidate set if it contains workforce vocabulary and receives a
screening score of at least 0.65. This procedure retains 173,273 excerpts and excludes
1,662,177 vocabulary-matched excerpts. To estimate how often relevant disruptions are
lost at this step, I classify a stratified sample of 250 excluded excerpts and
reweight the score-range-specific miss rates to the excluded population.

\begin{table}[!htbp]
\centering
\caption{Coverage Audit for Excerpts Excluded by the Candidate Screen}
\label{tab:app_gate}
\small
\begin{threeparttable}
\begin{tabular}{lrrrr}
\toprule
Screening-score range & Population weight & Classified & Disruptions identified & Miss rate\\
\midrule
\([0.55,0.65)\) & 0.0209 & 80 & 13 & 0.1625\\
\([0.45,0.55)\) & 0.0248 & 60 & 13 & 0.2167\\
\([0.30,0.45)\) & 0.0530 & 50 & 12 & 0.2400\\
\([0.00,0.30)\) & 0.9013 & 60 & 0 & 0.0000\\
\midrule
Population-reweighted & 1.0000 & 250 & 38 & 0.0215\\
\bottomrule
\end{tabular}
\begin{tablenotes}[flushleft]\footnotesize
\item Notes: The population-reweighted Wilson interval for the miss rate is
[0.0129, 0.0878]. Combining the point estimate with the estimated number of relevant
retained excerpts implies recall of approximately 0.73 for the candidate screen. The
separate firm-year validation exercise evaluates classification conditional on this
screen and therefore does not measure losses among excluded excerpts. Sentences
without workforce vocabulary are outside the construct's candidate set by definition.
\end{tablenotes}
\end{threeparttable}
\end{table}

\section{Rater Agreement and Coding-Rule Sensitivity}
\label{app:agreement}

The author--Opus result can be reconstructed from Table~\ref{tab:app_confusion}.
The difference in marginal positive rates is economically informative. I labeled
38.3\% of the sample positive, compared with 50.0\% for Opus. My case notes more often
required an unresolved adverse state and sometimes used a larger information window,
while the written broad-disruption criteria include costly adjustment and mitigated
events.

\begin{table}[!htbp]
\centering
\caption{Author--Opus Confusion Matrix on 120 Boundary-Case Excerpts}
\label{tab:app_confusion}
\small
\begin{threeparttable}
\begin{tabular}{lrrr}
\toprule
 & Opus positive & Opus negative & Author total\\
\midrule
Author positive & 32 & 14 & 46\\
Author negative & 28 & 46 & 74\\
\midrule
Opus total & 60 & 60 & 120\\
\bottomrule
\end{tabular}
\begin{tablenotes}[flushleft]\footnotesize
\item Notes: Agreement is 0.650 and Cohen's \(\kappa\) is 0.300, with bootstrap
95\% interval [0.134, 0.464]. The sample consists of excerpts identified by the broad
workforce-content screen and intentionally concentrates on cases near the construct
boundary.
\end{tablenotes}
\end{threeparttable}
\end{table}

\begin{table}[!htbp]
\centering
\caption{Agreement Under Alternative Applications of the Coding Rule}
\label{tab:app_protocol}
\scriptsize
\setlength{\tabcolsep}{3pt}
\begin{threeparttable}
\begin{tabular}{lrrrrrr}
\toprule
Interpretation & Agreement & \(\kappa\) & \shortstack{Both\\positive} &
\shortstack{Author positive\\only} & \shortstack{Opus positive\\only} &
\shortstack{Both\\negative}\\
\midrule
Original author classifications & 0.650 & 0.300 & 32 & 14 & 28 & 46\\
Conservative application of broad rule & 0.850 & 0.700 & 47 & 5 & 13 & 55\\
Central application of broad rule & 0.883 & 0.767 & 51 & 5 & 9 & 55\\
Inclusive application of broad rule & 0.908 & 0.817 & 54 & 5 & 6 & 55\\
\bottomrule
\end{tabular}
\begin{tablenotes}[flushleft]\footnotesize
\item Notes: A separate GPT-5.6 Sol review received both original classification
vectors and applied the written broad-disruption rule to the 42 disagreements; the 78
agreements were held fixed. The resulting rows measure sensitivity to alternative
applications of the coding rule. They are not independent third-rater estimates and
do not replace the original author--Opus comparison.
\end{tablenotes}
\end{threeparttable}
\end{table}

\section{Alternative Classification Rules}
\label{app:ambiguity}

Table~\ref{tab:app_variant_size} shows that the alternative rules are economically
meaningful rather than cosmetic. The inclusive rule and the rule requiring agreement
of at least two classifiers identify more than 60\% additional excerpts, while the
high-confidence rule and the rule requiring agreement of all three classifiers
identify roughly one-third fewer. Even so, firm-year rankings remain close, and the
content-based exclusions barely change the annual ordering.

\begin{table}[!htbp]
\centering
\caption{Firm-Year Measure Stability Across Classification Rules}
\label{tab:app_variant_size}
\footnotesize
\begin{threeparttable}
\begin{tabularx}{\linewidth}{Xrrrr}
\toprule
Definition & \shortstack{Disruption\\excerpts} & \shortstack{Relative to\\baseline} &
Spearman & \shortstack{Top-decile\\Jaccard}\\
\midrule
Baseline & 45,232 & 1.000 & 1.000 & 1.000\\
Inclusive threshold & 78,003 & 1.725 & 0.889 & 0.670\\
High-confidence threshold & 30,294 & 0.670 & 0.914 & 0.726\\
At least two of three classifiers & 73,007 & 1.614 & 0.899 & 0.684\\
All three classifiers & 28,529 & 0.631 & 0.900 & 0.704\\
Excluding explicit labor-shortage passages & 42,792 & 0.946 & 0.989 & 0.905\\
Excluding leadership and succession passages & 43,121 & 0.953 & 0.976 & 0.911\\
\bottomrule
\end{tabularx}
\begin{tablenotes}[flushleft]\footnotesize
\item Notes: Spearman correlations and within-year top-decile Jaccard overlaps compare
each firm-year score with the baseline in the main FY2006--FY2024 sample.
\end{tablenotes}
\end{threeparttable}
\end{table}

\begin{landscape}
\begin{table}[!htbp]
\centering
\caption{Annual Economic Coefficients Across Classification Rules}
\label{tab:app_economic_grid}
\footnotesize
\begin{threeparttable}
\begin{tabular}{lrrrr}
\toprule
Definition & Idiosyncratic volatility & Downside deviation & Worst monthly return &
Market beta\\
\midrule
Baseline & 0.005454 (3.43) & 0.005840 (3.92) & -0.004592 (-3.75) & -0.001396 (-0.37)\\
Inclusive threshold & 0.005559 (3.32) & 0.006242 (4.00) & -0.004426 (-3.48) & -0.002296 (-0.56)\\
High-confidence threshold & 0.006427 (4.19) & 0.006183 (4.25) & -0.004900 (-4.12) & 0.001200 (0.33)\\
At least two of three classifiers & 0.005616 (3.35) & 0.006072 (3.89) & -0.004386 (-3.46) & -0.002365 (-0.58)\\
All three classifiers & 0.006540 (4.24) & 0.006097 (4.18) & -0.004804 (-4.00) & 0.001612 (0.45)\\
Excluding explicit labor-shortage passages & 0.005698 (3.50) & 0.006034 (4.05) & -0.004818 (-3.94) & -0.000505 (-0.13)\\
Excluding leadership and succession passages & 0.005031 (3.16) & 0.005609 (3.73) & -0.004467 (-3.64) & -0.000258 (-0.07)\\
\bottomrule
\end{tabular}
\begin{tablenotes}[flushleft]\footnotesize
\item Notes: Entries are coefficients per within-sample standard deviation with
firm-clustered \(t\)-statistics in parentheses. Every model includes broad workforce
exposure, firm controls, and firm and fiscal-year fixed effects. Samples are held
fixed across variants within each outcome.
\end{tablenotes}
\end{threeparttable}
\end{table}
\end{landscape}

\begin{figure}[!htbp]
\centering
\includegraphics[width=0.95\linewidth]{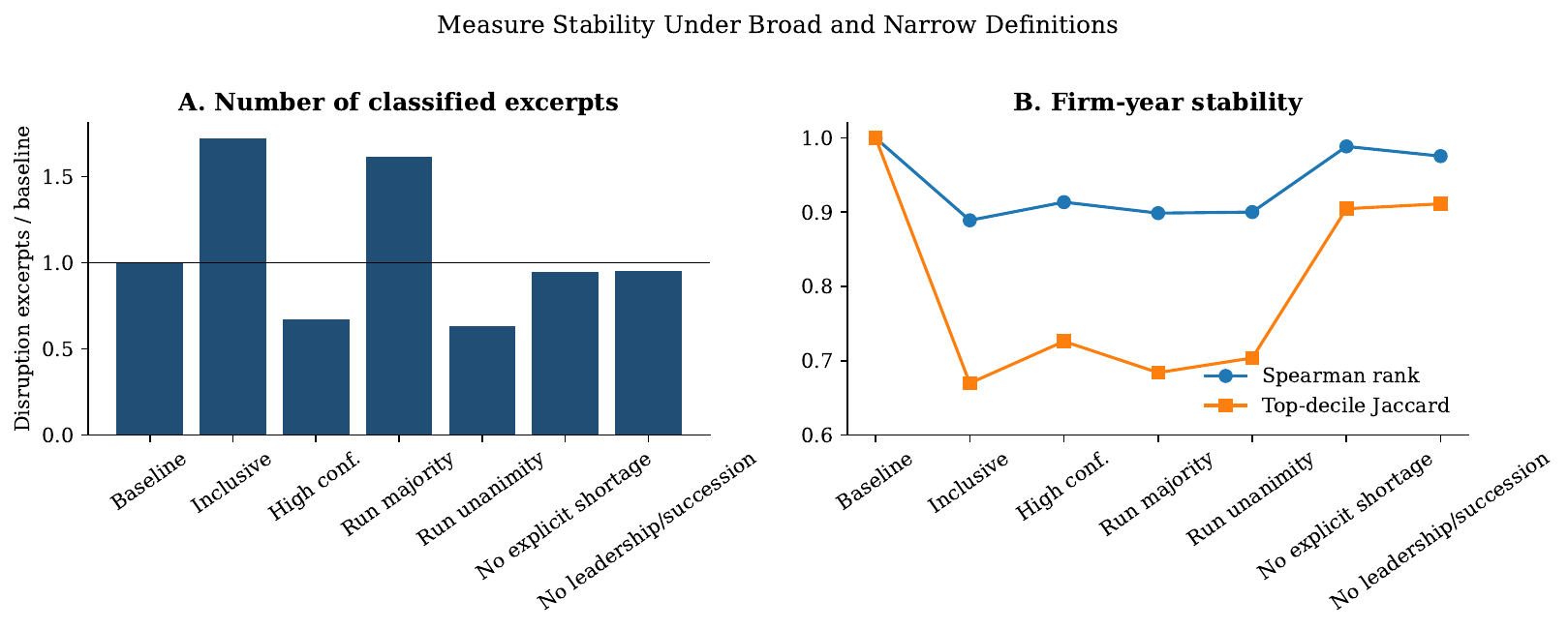}
\caption{Score Size and Rank Stability Under Alternative Classification Rules}
\label{fig:app_measure_stability}
\end{figure}

\begin{table}[!htbp]
\centering
\caption{Fama--French Industries Classified as High Labor Intensity}
\label{tab:app_labor_industries}
\small
\begin{threeparttable}
\begin{tabular}{rlrl}
\toprule
FF48 & Industry & FF48 & Industry\\
\midrule
6  & Recreation products & 22 & Electrical equipment\\
7  & Entertainment & 23 & Automobiles and trucks\\
8  & Printing and publishing & 24 & Aircraft\\
9  & Consumer goods & 25 & Shipbuilding and railroad equipment\\
10 & Apparel & 26 & Defense\\
11 & Healthcare & 33 & Personal services\\
15 & Rubber and plastic products & 35 & Computers\\
16 & Textiles & 38 & Business supplies\\
17 & Construction materials & 39 & Shipping containers\\
20 & Fabricated products & 41 & Wholesale\\
21 & Machinery & 42 & Retail\\
   & & 43 & Restaurants, hotels, and motels\\
   & & 46 & Real estate\\
\bottomrule
\end{tabular}
\begin{tablenotes}[flushleft]\footnotesize
\item Notes: For each Fama--French 48 industry, I compute the median ratio of
Compustat employment to assets among call-covered firm-years from 2006 through 2024.
The table lists the 24 industries whose industry-level median exceeds the median
across the 48 industries.
\end{tablenotes}
\end{threeparttable}
\end{table}

\section{Additional Annual Results}
\label{app:annual}

Appendix Table~\ref{tab:app_annual_robustness} reports the specification sensitivities
summarized in Section~\ref{sec:baseline}. Appendix
Table~\ref{tab:app_long_horizon} reports the longer-horizon outcomes.

\begin{table}[!htbp]
\centering
\caption{Annual Idiosyncratic-Volatility Specification Sensitivity}
\label{tab:app_annual_robustness}
\footnotesize
\begin{threeparttable}
\begin{tabularx}{\linewidth}{Xrrr}
\toprule
Specification & Coefficient & $t$ & $N$\\
\midrule
\multicolumn{4}{l}{\textit{Panel A. Samples, transformations, and inference}}\\
Baseline, firm-clustered & 0.0055*** & 3.43 & 12,114\\
Two-way firm and fiscal-year clustering & 0.0055*** & 2.86 & 12,114\\
Exclude 2020--2021 & 0.0053*** & 2.68 & 7,358\\
FY2006--FY2014 & 0.0094** & 2.46 & 3,036\\
FY2015--FY2024 & 0.0033* & 1.95 & 9,052\\
Inverse-hyperbolic-sine score & 0.0056*** & 3.59 & 12,114\\
Outcome winsorized at 0.5 and 99.5 percent & 0.0058*** & 3.42 & 12,114\\
\midrule
\multicolumn{4}{l}{\textit{Panel B. Alternative score constructions}}\\
Inclusive maximum-$F_1$ threshold & 0.0056*** & 3.32 & 12,114\\
Disruption share among candidate excerpts & 0.0060*** & 3.96 & 12,114\\
Continuous score among positive firm-years & 0.0045** & 2.54 & 7,049\\
\bottomrule
\end{tabularx}
\begin{tablenotes}[flushleft]\footnotesize
\item Notes: The dependent variable is annualized idiosyncratic volatility. Except where stated otherwise, models include firm and fiscal-year fixed effects, broad workforce discussion, and standard firm controls, with firm-clustered standard errors. The score is standardized within each estimation sample. The first row reproduces the principal estimate.
\end{tablenotes}
\end{threeparttable}
\end{table}

\begin{table}[!htbp]
\centering
\caption{Longer-Horizon Risk, Operating, and Return Outcomes}
\label{tab:app_long_horizon}
\footnotesize
\begin{threeparttable}
\begin{tabularx}{\linewidth}{Xrrr}
\toprule
Outcome & Coefficient & $t$ & $N$\\
\midrule
Next-year idiosyncratic volatility & 0.0017 & 0.93 & 8,741\\
Next-year total volatility & -0.0002 & -0.10 & 9,032\\
Employment growth & 0.0031 & 0.65 & 11,743\\
Absolute employment growth & 0.0002 & 0.10 & 8,702\\
Sales growth & 0.0033 & 0.45 & 11,853\\
Absolute sales growth & -0.0176* & -1.79 & 8,743\\
Capital expenditure/assets & -0.0001 & -0.37 & 11,847\\
ROA & -0.0009 & -1.14 & 11,878\\
Twelve-month-ahead return & -0.0007 & -0.13 & 12,520\\
\bottomrule
\end{tabularx}
\begin{tablenotes}[flushleft]\footnotesize
\item Notes: Entries are coefficients per within-sample standard deviation of the disruption score, with firm-clustered \(t\)-statistics. Models include firm and fiscal-year fixed effects, broad workforce discussion, and standard firm controls. Outcome construction follows Section~\ref{sec:data}. The estimates do not reveal a common directional relation between the score and average next-year operating outcomes; the negative coefficient for absolute sales growth is marginal at the 10\% level.
\end{tablenotes}
\end{threeparttable}
\end{table}

\section{Comparison with Labor-Shortage Exposure}
\label{app:hhq}

I use the labor-shortage exposure released by \textcite{harford2026labor}, hereafter
HHQ. For descriptive overlap, I aggregate this paper's measure to calendar year and
match the two measures by firm-year. For the joint annual regressions, I match the HHQ
calendar-year observation to this paper's fiscal-year observation carrying the same
numerical year; the underlying intervals coincide exactly only for December-year-end
firms. These comparisons distinguish the constructs but do not reproduce HHQ's
original outcome specifications.

\begin{table}[!htbp]
\centering
\caption{Comparison with the Harford--He--Qiu (HHQ) Labor-Shortage Measure}
\label{tab:hhq_overlap}
\footnotesize
\begin{threeparttable}
\begin{tabularx}{\linewidth}{>{\raggedright\arraybackslash}X >{\centering\arraybackslash}p{2.5cm} >{\centering\arraybackslash}p{4.2cm}}
\toprule
Quantity & All disruption passages & HC measure excluding excerpts with explicit labor-shortage phrases\\
\midrule
\multicolumn{3}{l}{\textit{Panel A. Prevalence of HHQ labor-shortage exposure}}\\
Call-quarter positive share & 0.131 & --\\
Firm-year positive share & 0.297 & --\\
\midrule
\multicolumn{3}{l}{\textit{Panel B. Common one-call firm-calendar-quarters ($N=25,307$)}}\\
HC-positive share & 0.338 & 0.333\\
HHQ-positive share & 0.184 & 0.184\\
$P(HC^+\mid HHQ^+)$ & 0.622 & 0.596\\
$P(HHQ^+\mid HC^+)$ & 0.338 & 0.329\\
Positive-set Jaccard & 0.281 & 0.269\\
Pearson correlation, continuous scores & 0.476 & 0.406\\
\midrule
\multicolumn{3}{l}{\textit{Panel C. Common firm-calendar-years ($N=8,047$)}}\\
$P(HC^+\mid HHQ^+)$ & 0.771 & 0.759\\
$P(HHQ^+\mid HC^+)$ & 0.506 & 0.504\\
Positive-set Jaccard & 0.440 & 0.434\\
Pearson correlation, continuous scores & 0.531 & 0.474\\
\bottomrule
\end{tabularx}
\begin{tablenotes}[flushleft]\footnotesize
\item Notes: HHQ denotes the labor-shortage exposure measure of Harford, He, and Qiu; HC denotes this paper's human-capital disruption measure. Panel A reports prevalence in HHQ's 138,159 call-quarter observations and 39,291 firm-year observations. Panel B contains common firm-calendar-quarters in which this paper's transcript sample has exactly one call. The HHQ data do not provide call dates or transcript text, so the observations cannot be linked transcript by transcript and the two studies' transcript collections may differ. Panel C aligns observations by firm and calendar year for descriptive comparison. This calendar-year alignment differs from the fiscal-year timing of the paper's annual regressions. The final column removes disruption passages containing explicit shortage phrases before aggregation.
\end{tablenotes}
\end{threeparttable}
\end{table}

\begin{figure}[!htbp]
\centering
\includegraphics[width=0.86\linewidth]{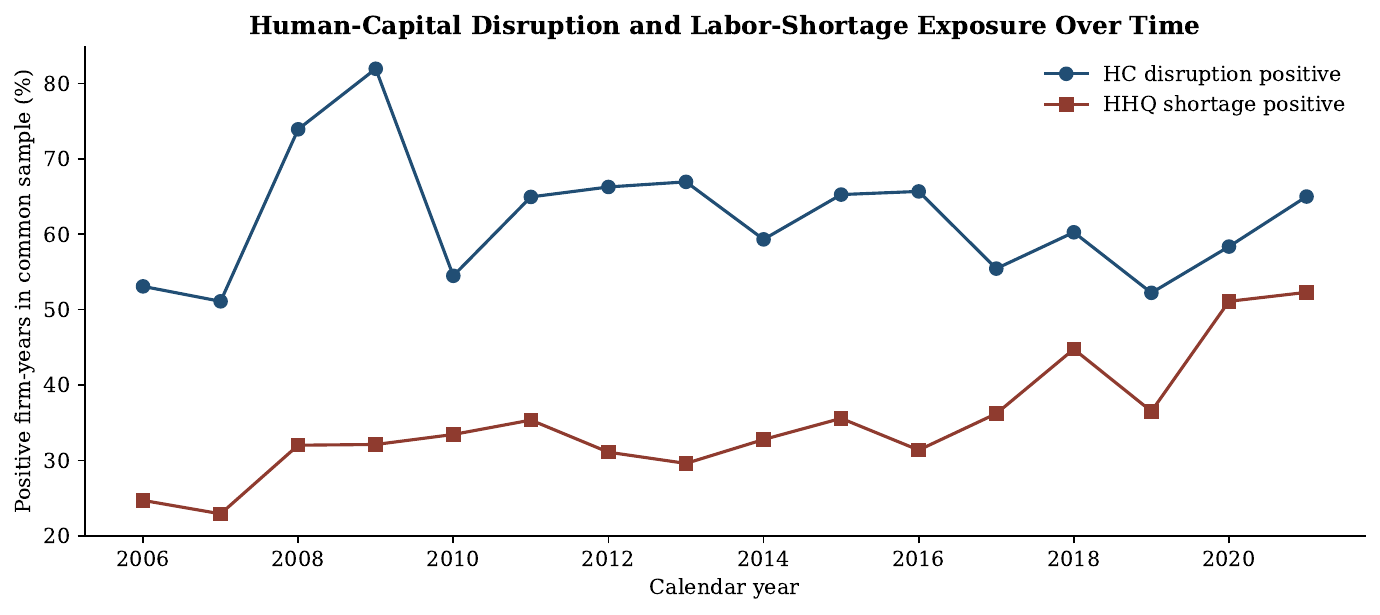}
\caption{Human-Capital Disruption and Labor-Shortage Exposure Over Time}
\label{fig:app_hhq_time}
\begin{minipage}{0.90\linewidth}
\footnotesize
\textit{Notes:} The figure reports annual positive shares on the common
gvkey--calendar-year sample from 2006 through 2021. It does not compare raw
within-document frequencies because the two measures use different denominators and
transcript collections.
\end{minipage}
\end{figure}

Let \(HHQ^+\) denote a positive value of their binary labor-shortage indicator.
Across the seven disruption definitions, \(P(HC^+\mid HHQ^+)\) ranges from 66.5\%
when all three classifiers must agree to 88.3\% under the inclusive rule. Excluding
explicit labor-shortage passages leaves the conditional probability at 75.9\%. The
asymmetric relation is therefore not generated solely by one threshold or by direct
shortage keywords.

\subsection{Generic negative and uncertain language}

I score each transcript in this paper using the March 2026 Loughran--McDonald Master
Dictionary. The benchmark variables are negative and uncertainty words per 10,000
transcript words. They use the full transcript, are not estimated from the
human-capital labels, and enter the annual and call-level specifications in
standardized form.

The matched-transcript comparison shows that generic negative language captures part,
but not all, of the same variation. In the full 42-day sample, the human-capital
coefficient falls from 0.00491 (\(t=2.09\)) to 0.00332 (\(t=1.59\)) when negative
and uncertainty frequencies are added. Among firm-quarters containing exactly one
call in this corpus, the coefficient is 0.00653 (\(t=2.17\)) when the HHQ indicator
and both tone measures enter jointly. In annual regressions, the coefficient is
0.00651 (\(t=2.97\)), and it is 0.00683 (\(t=3.28\)) after explicit
labor-shortage passages are excluded. The call-level attenuation indicates overlap
with generic negativity, while the annual and common-quarter estimates retain
variation distinct from generic tone and the HHQ shortage measure.

\section{Additional Call-Timed Results}
\label{app:call}

Table~\ref{tab:app_dynamic_summary} summarizes the risk state before the call and the
risk realizations that follow it. The grouped pre-call coefficients describe the
state that precedes disclosure. The post-call estimates condition on the nearest
pre-call block, while the change estimates compare the post-call and pre-call states.

\begin{table}[!htbp]
\centering
\caption{Dynamic Risk-Path Summary}
\label{tab:app_dynamic_summary}
\footnotesize
\begin{threeparttable}
\begin{tabular}{lrr}
\toprule
Relative period or test & Idiosyncratic volatility & Downside deviation\\
\midrule
Pre-call days \(-169\) to \(-107\) & 0.63\% (2.25) & 0.71\% (2.47)\\
Pre-call days \(-106\) to \(-44\) & 0.69\% (2.49) & 0.80\% (2.76)\\
Pre-call days \(-43\) to \(-2\) & 0.85\% (3.36) & 0.87\% (3.44)\\
Post-call days \(+2\) to \(+22\), conditional & 0.54\% (2.14) & 0.36\% (1.38)\\
Post-call days \(+23\) to \(+43\), conditional & 0.56\% (2.09) & 0.62\% (2.30)\\
Unconditional post-minus-pre change &
\(0.12\%\;[-0.27,\,0.50]\;(0.59)\) &
\(-0.25\%\;[-0.74,\,0.23]\;(-1.03)\)\\
\bottomrule
\end{tabular}
\begin{tablenotes}[flushleft]\footnotesize
\item Notes: For each pre-call interval, the dependent variable averages realized
risk across the constituent nonoverlapping 21-trading-day blocks before estimation.
State and conditional-post entries are approximate percentage changes per score
standard deviation, with \(t\)-statistics in parentheses. Change entries report the
coefficient, 95\% confidence interval, and \(t\)-statistic. Because both change
intervals include zero, the evidence supports a persistent risk state around the call
rather than a discrete change at disclosure.
\end{tablenotes}
\end{threeparttable}
\end{table}

\begin{landscape}
\begin{table}[!htbp]
\centering
\caption{Secondary Call-Timed Outcomes and Horizons}
\label{tab:app_call_secondary}
\footnotesize
\begin{threeparttable}
\begin{tabular}{p{4.2cm}p{3.8cm}p{5.0cm}rrrr}
\toprule
Horizon and sample & Outcome & Score definition & Association per SD & $t$ & $N$ & Firms\\
\midrule
42 days; All calls & Tail-loss magnitude & Baseline & 0.60\% & 1.23 & 41,787 & 2,527\\
42 days; All calls & Market beta & Baseline & 0.0010 & 0.32 & 41,787 & 2,527\\
42 days; Before next observed call & Tail-loss magnitude & Baseline & 0.79\%* & 1.68 & 38,357 & 2,389\\
42 days; Before next observed call & Market beta & Baseline & 0.0013 & 0.46 & 38,357 & 2,389\\
42 days; Before next observed call & Idiosyncratic volatility & Excluding leadership/succession passages & 0.46\%* & 1.80 & 38,357 & 2,389\\
42 days; Before next observed call & Downside deviation & Excluding leadership/succession passages & 0.41\% & 1.60 & 38,357 & 2,389\\
\midrule
63 days; All calls & Idiosyncratic volatility & Baseline & 0.69\%*** & 3.22 & 41,777 & 2,527\\
63 days; All calls & Downside deviation & Baseline & 0.50\%** & 2.17 & 41,777 & 2,527\\
63 days; All calls & Tail-loss magnitude & Baseline & 0.88\%*** & 2.65 & 41,777 & 2,527\\
63 days; All calls & Market beta & Baseline & 0.0006 & 0.20 & 41,777 & 2,527\\
63 days; Before next observed call & Idiosyncratic volatility & Baseline & 0.79\%** & 2.16 & 11,191 & 1,671\\
63 days; Before next observed call & Downside deviation & Baseline & 0.59\% & 1.46 & 11,191 & 1,671\\
63 days; Before next observed call & Tail-loss magnitude & Baseline & 0.93\%* & 1.65 & 11,191 & 1,671\\
63 days; Before next observed call & Market beta & Baseline & 0.0066 & 1.37 & 11,191 & 1,671\\
\bottomrule
\end{tabular}
\begin{tablenotes}[flushleft]\footnotesize
\item Notes: The specifications match Table~\ref{tab:call_timed}. Positive risk outcomes are logged and reported as coefficient-implied percentage differences; market beta is in levels. ``Before next observed call'' requires an observed successor call and a complete outcome window that ends before it. Among calls with an observed successor, 30.3\% of valid 63-day windows satisfy this restriction. Tail-loss magnitude is the negative of the most negative rolling 21-trading-day sum of Fama--French three-factor residual returns within the outcome window.
\end{tablenotes}
\end{threeparttable}
\end{table}
\end{landscape}

The 63-day idiosyncratic-volatility coefficient remains positive after a 99th-percentile
winsorization of the measure, an inverse-hyperbolic-sine transformation, removal of the
top 1\% of score values, exclusion of 2020--2021, restriction to the last call in
each firm-quarter, firm-only clustering, and industry-by-quarter absorption. An
approximate top-0.5\% Cook-distance deletion leaves \(t=2.79\). Downside and tail
outcomes are less stable across specifications: their coefficients attenuate over the
cumulative 42-day window, with industry-by-quarter fixed effects, and under more
inclusive classification rules. The most consistent call-level evidence is therefore
the association with idiosyncratic volatility.

\section{Executive Outcomes and Public-Information Screens}
\label{app:turnover}

The annual roster outcomes and the exact CEO-office-exit outcome have different units
and meanings. Roster absence means that a person is not in the exact next-year named
executive group; it need not indicate that the person left the company. The call-level
CEO event uses the effective date of leaving the CEO office, strictly within a
call-to-next-call interval. Neither outcome is an announcement date.

\begin{table}[!htbp]
\centering
\caption{CEO-Office-Exit Influence Results}
\label{tab:app_ceo_diagnostics}
\small
\begin{threeparttable}
\begin{tabular}{lrr}
\toprule
Specification & Coefficient per score SD & \(t\)-statistic\\
\midrule
Fully controlled baseline & 0.391 pp & 2.53\\
Drop most influential firm & 0.339 pp & 2.27\\
Drop five most influential firms & 0.308 pp & 2.10\\
Drop approximate top 0.5\% Cook distances & 0.147 pp & 1.37\\
\bottomrule
\end{tabular}
\begin{tablenotes}[flushleft]\footnotesize
\item Notes: Observation-level Cook trimming removes a disproportionate number of
rare positive events and is not the preferred specification.
\end{tablenotes}
\end{threeparttable}
\end{table}

To assess whether the CEO-office-exit association reflects succession information
already public before the call, I link the 715 exits to SEC filings. The search covers
all 3,270 identified pre-call event--document pairs, representing 3,135 distinct
primary documents, and all 66 same-day pairs associated with these exits. Eleven exits
have no Item 5.02 filing candidate within the prior 548 days. The narrower exclusion
removes 185 observations for which directional incumbent-transition language appears
near a date matching the Execucomp effective date. A broader exclusion removes 401
observations after expanding both the language and same-day criteria. Both
specifications therefore exclude observations with affirmative evidence of a prior
public disclosure. Because the procedure is designed to identify such evidence rather
than prove its absence, retained observations are not described as unannounced exits.

\section{Variable Timing and Definitions}
\label{app:reproducibility}

\begin{table}[!htbp]
\centering
\caption{Principal Variables, Timing, and Interpretation}
\label{tab:app_variables}
\footnotesize
\setlength{\tabcolsep}{4pt}
\renewcommand{\arraystretch}{0.96}
\begin{threeparttable}
\begin{tabularx}{\linewidth}{p{3.3cm}p{4.2cm}X}
\toprule
Variable & Timing and construction & Interpretation\\
\midrule
Annual HC disruption & Classified disruption excerpts between consecutive fiscal endpoints,
divided by transcript words & Fiscal-year disclosure
frequency; contemporaneous in annual tests\\
Broad workforce discussion & Workforce-vocabulary sentences divided by transcript
words & Workforce discussion control defined before disruption classification\\
Annual idiosyncratic volatility & FF3 residual volatility over the 365 calendar days
ending at fiscal year-end & Firm-specific return dispersion\\
Downside deviation & Annualized dispersion of negative firm-specific returns in the
fiscal window & Adverse side of the firm-specific distribution\\
Worst monthly return & Minimum monthly return in the fiscal window & Tail-oriented
realized return outcome\\
Post-call risk & Trading day \(+2\) through the selected 21-, 42-, or
63-trading-day horizon; immediate response days omitted & Subsequent risk level,
conditional on matched pre-call risk\\
Executive-roster exit & Absence from exact next fiscal-year Execucomp named roster &
Roster churn, not confirmed company departure\\
CEO-office exit & \texttt{LEFTOFC} strictly after the call and before the next call &
Effective departure from CEO office, not announcement or forced-turnover status\\
\bottomrule
\end{tabularx}
\begin{tablenotes}[flushleft]\footnotesize
\item Notes: Firm-year standardizations are outcome-sample specific. Call-level
accounting controls use the latest fiscal-Q4 report date strictly before the call.
\end{tablenotes}
\end{threeparttable}
\end{table}

\end{document}